\documentclass[iop]{emulateapj}
\usepackage{apjfonts}
\usepackage{natbib}
\usepackage{epstopdf}
\usepackage{graphicx}
\def\familyone{\mbox{family-1\ }}

\def\familytwo{\mbox{family-2\ }}

\begin{document}

\title{GRB 140619B: a short GRB from a binary neutron star merger leading to black hole formation}

\author{R. Ruffini\altaffilmark{1,2,3,4}, M. Muccino\altaffilmark{1,2}, M. Kovacevic\altaffilmark{1,3}, F.~G. Oliveira\altaffilmark{1,3}, J.~A. Rueda\altaffilmark{1,2,4}, C.~L. Bianco\altaffilmark{1,2},\\ M. Enderli\altaffilmark{1,3}, A.~V. Penacchioni\altaffilmark{4,5}, G.~B. Pisani\altaffilmark{1,3}, Y.~Wang\altaffilmark{1,2}, E. Zaninoni\altaffilmark{4}}

\altaffiltext{1}{Dip. di Fisica and ICRA, Sapienza Universit\`a di Roma, Piazzale Aldo Moro 5, I-00185 Rome, Italy.}
\altaffiltext{2}{ICRANet, Piazza della Repubblica 10, I-65122 Pescara, Italy.}
\altaffiltext{3}{Universit\'e de Nice Sophia Antipolis, CEDEX 2, Grand Ch\^{a}teau Parc Valrose, Nice, France.}
\altaffiltext{4}{ICRANet-Rio, Centro Brasileiro de Pesquisas Fisicas, Rua Dr. Xavier Sigaud 150, Rio de Janeiro, RJ, 22290-180, Brazil.}
\altaffiltext{5}{Instituto Nacional de Pesquisas Espaciais, Av. dos Astronautas, 1758, S\~ao Jos\'e dos Campos, SP, 12227-010, Brazil.}

\shorttitle{GRB 140619B}

\shortauthors{Ruffini et al.}

\begin{abstract}
We show the existence of two families of short GRBs, both originating from the merger of binary neutron stars (NSs): \familyone with $E_{iso}<10^{52}$ erg, leading to a massive NS  as the merged core, and \familytwo with $E_{iso}>10^{52}$ erg, leading to a black hole (BH).
Following the identification of the prototype GRB 090227B, we present the details of a new example of \familytwo short burst: GRB 140619B. 
From the spectral analysis of the early $\sim0.2$ s, we infer an observed temperature $kT =(324\pm33)$ keV of the $e^+e^-$-plasma at transparency (P-GRB), a theoretically derived redshift $z=2.67\pm0.37$, a total burst energy $E^{tot}_{e^+e^-}=(6.03\pm0.79)\times10^{52}$ erg, a rest-frame peak energy $E_{p,i}=4.7$ MeV, and a baryon load $B=(5.52\pm0.73)\times10^{-5}$.
We also estimate the corresponding emission of gravitational waves.
Two additional examples of \familytwo short bursts are identified: GRB 081024B and GRB 090510, remarkable for its well determined cosmological distance.
We show that marked differences exist in the nature of the afterglows of these two families of short bursts: \familytwo bursts, leading to BH formation, consistently exhibit high energy emission following the P-GRB emission; \familyone bursts, leading to the formation of a massive NS, should never exhibit high energy emission. We also show that both the families fulfill an $E_{p,i}$--$E_{iso}$ relation with slope $\gamma=0.59\pm0.07$ and a normalization constant incompatible with the one for long GRBs.
The observed rate of such \familytwo events is $\rho_0=\left(2.1^{+2.8}_{-1.4}\right)\times10^{-4}$Gpc$^{-3}$yr$^{-1}$.
\end{abstract}


\email{ruffini@icra.it}

\maketitle

\section{Introduction}\label{sec:1}

The phenomenological classification of gamma-ray bursts (GRBs) based on their prompt emission observed $T_{90}$ durations defines  ``long'' and ``short'' bursts which are, respectively, longer or shorter than $T_{90}=2$ s \citep{Klebesadel1992,Dezalay1992,Koveliotou1993,Tavani1998}.
Short GRBs have been often indicated as originating from binary neutron star (NS) mergers \citep[see, e.g.,][]{Goodman1986,Paczynski1986,Eichler1989,Narayan1991,MeszarosRees1997_b,Rosswog2003,Lee2004,2014ARA&A..52...43B}.

An ample literature exists of short GRBs with a measured redshift, isotropic burst energy $E_{iso}<10^{52}$ erg and rest-frame spectral peak energy $E_{p,i}<2$ MeV \citep[see, e.g.,][and references therein]{2014ARA&A..52...43B}.
Thanks to extensive data provided by the \textit{Swift}-XRT instrument \citep{Burrows2005}, it is possible to observe the long lasting X-ray afterglow of these short bursts to identify their host galaxies and to compute their cosmological redshifts.
They have been observed in both early- and late-type galaxies with older stellar population ages \citep[see, e.g.,][for details]{2014ARA&A..52...43B}, and at systematically larger radial offsets from their host galaxies than long GRBs \citep{Sahu1997,vanParadijs1997,Bloom2006,Troja2008,Fong2010,Berger2011,Kopaz2012}.
None of these afterglows appears to have the specific power law signature  in the X-ray luminosity when computed in the source rest-frame, as found in some long GRBs \citep[see, e.g.,][]{Ruffini2014}.

In the meantime, considerable progress has been obtained in the theoretical understanding of the equilibrium configuration of NSs, in their mass-radius relation  (see Fig.~\ref{fig:mass_rad} in Sec.~\ref{sec:NS}), and especially in the theoretical determination of the value of the NS critical mass for gravitational collapse $M_{crit}^{NS}$ \citep{Rotondo2011,Rueda2011,Belvedere}. This has led to a theoretical value $M_{crit}^{NS}=2.67$ M$_\odot$ \citep{Belvedere}. Particularly relevant to this determination has been the conceptual change of paradigm of imposing  global charge neutrality  \citep{Belvedere} instead of the traditional local charge neutrality (LCN) still applied in the current literature \citep[see, e.g.,][and references therein]{2007ASSL..326.....H}.

Similarly, noteworthy progress has been achieved in the determination of the masses of galactic binary pulsars.
Of the greatest relevance has been the direct observation of NS masses larger than $2$ M$_\odot$ (see \citealp{2013Sci...340..448A} and Sec.~\ref{sec:NS}).
In the majority of the observed cases of binary NSs the sum of the NS masses, $M_1+M_2$, is indeed smaller than $M_{crit}^{NS}$ and, given the above determination of the NS critical mass, their coalescence will never lead to a BH formation (see Fig.~\ref{fig:dns} in Sec.~\ref{sec:NS}). 
This of course offers a clear challenge to the traditional assumption that all short GRBs originate from BH formation \citep[see, e.g.,][and references therein]{2014ARA&A..52...43B}.

Motivated by the above considerations, we propose in this article the existence of two families of short GRBs, both originating from NS mergers: the difference between these two families depends on whether the total mass of the merged core is smaller or larger than $M_{crit}^{NS}$. 
We assume that \familyone coincides with the above mentioned less energetic short GRBs with $E_{iso}<10^{52}$ erg and the coalescence of the merging NSs leads to a massive NS as the merged core. 
We assume that \familytwo short bursts with $E_{iso}>10^{52}$ erg originate from a merger process leading to a BH as the merged core. 
The presence of the BH allows us to address the GRB nature within the fireshell model \citep{Ruffini2001c,Ruffini2001,Ruffini2001a} leading to specific signatures in the luminosity, spectra and time variability observed in two very different components: the proper-GRB (P-GRB) and the prompt emission (see Sec.~\ref{sec:2}). 
The prototype is GRB 090227B, which we already analyzed within the fireshell model in \citet{Muccino2013}.
We also assume that the BH gives rise to the short-lived ($\lesssim10^2$ s in the observer frame) and very energetic GeV emission which has been found to be present in all these \familytwo short GRBs, when \textit{Fermi}-LAT data are available.
This article is mainly dedicated to giving the theoretical predictions and the observational diagnostics to support the above picture.

In Sec.~\ref{GRB090227B} we recall the results obtained in the case of the prototype of \familytwo short GRBs: GRB 090227B \citep{Muccino2013}.
The analysis of its P-GRB emission led to a particularly low value of the baryon load, $B\sim10^{-5}$, as well as to the prediction of the distance corresponding to a redshift $z=1.61$, and consequently to $E^{tot}_{e^+e^-}=2.83\times10^{53}$ erg.
From the analysis of the spectrum and the light curve of the prompt emission we inferred an average circumburst medium (CBM) density $\langle n_{CBM}\rangle\sim10^{-5}$ cm$^{-3}$ typical of galactic halos of GRB host galaxies.

In Sec.~\ref{sec:3} we summarize the observations of a second example of such \familytwo short bursts, GRB 140619B, and our data analysis from $8$ keV up to $100$ GeV. We also point out the lack of any observed X-ray afterglow following the prompt emission \citep{GCN16424}.

\begin{figure}
\centering
\includegraphics[width=\hsize,clip]{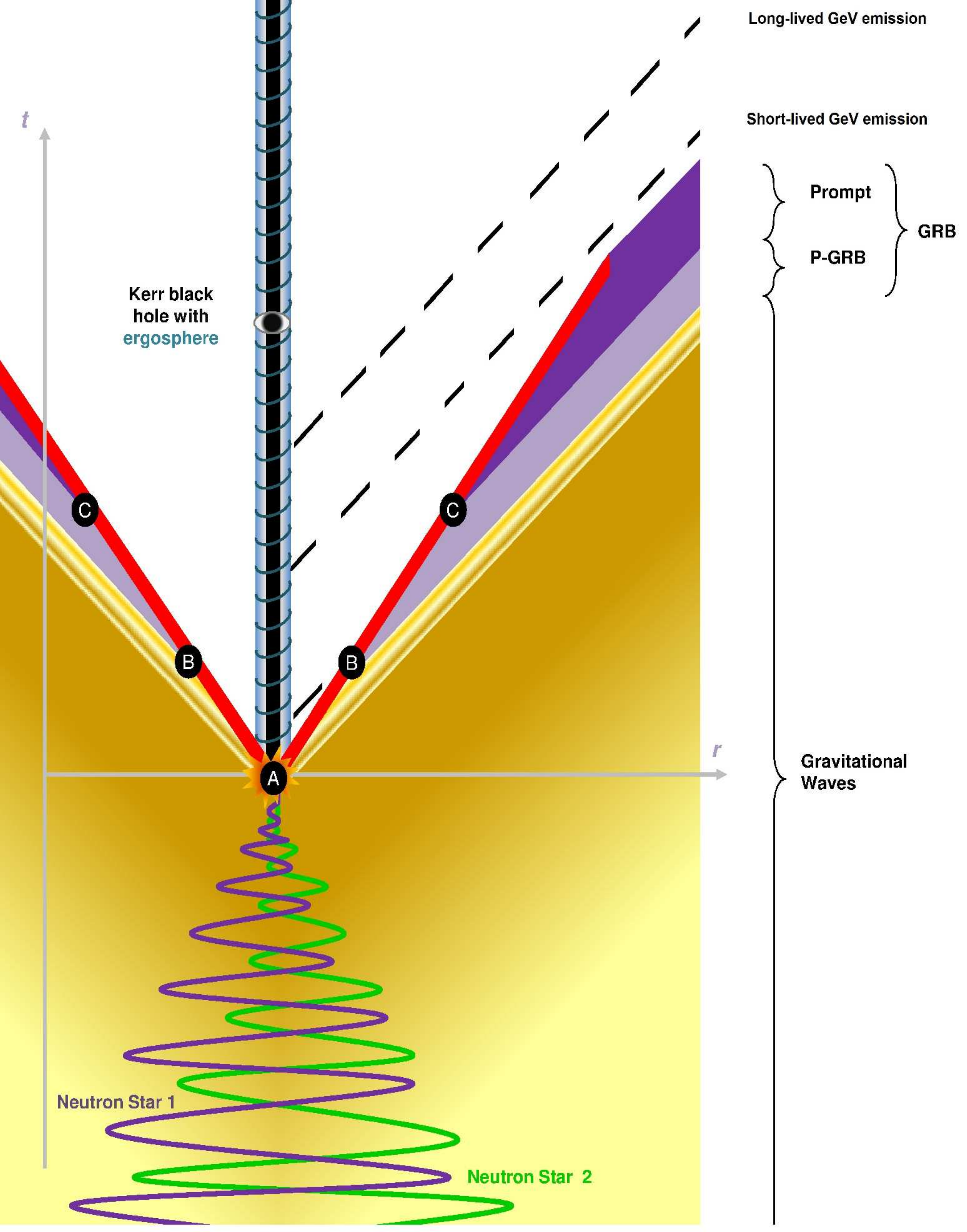}
\caption{The space-time diagram of \familytwo short GRBs. The orbital separation between the two NSs decreases due to the emission of GWs, until the merging occurs and a \familytwo short GRB is emitted. Following the fireshell model (see Sec.~\ref{sec:2}): A) vacuum polarization occurs while the event horizon is formed and a fireshell of $e^+e^-$ plasma self-accelerates radially outwards; B) the fireshell, after engulfing the baryons, keeps self-accelerating and reaches the transparency when the P-GRB is emitted; C) the accelerated baryons interact with the local CBM giving rise to the prompt emission. The remnant of the merger is a Kerr BH. The accretion of a small (large) amount of orbiting matter onto the BH can lead to the short lived but very energetic $0.1$--$100$ GeV emission observed in GRB 081024B, GRB 090510 and GRB 140619B. The absence of such an emission in GRB 090227B is due to the absence of observations of \textit{Fermi}-LAT.}
\label{fig:diagram}
\end{figure}

In Sec.~\ref{sec:4} we address GRB 140619B within the fireshell model and compare and contrast the results with those of the prototype, GRB 090227B \citep{Muccino2013}.
In Sec.~\ref{sec:4.1}, from the fireshell equations of motion, we theoretically estimate and predict the value of the redshift of the source, $z=2.67\pm0.37$.
Consequently, we derive the burst energy $E_{iso}>10^{52}$ erg and the value of the baryon load $B\sim10^{-5}$.
In Sec.~\ref{sec:4.2} we infer an average density of the CBM $\langle n_{CBM}\rangle\sim10^{-5}$ cm$^{-3}$ from fitting the prompt emission light curve and spectra. This parameter is typical of the galactic halo environment and further confirms a NS--NS merger as the progenitor for GRB 140619B (see Sec.~\ref{sec:4.3} and Fig.~\ref{fig:diagram}).

In Sec.~\ref{GWs} we discuss the possibility for Advanced LIGO to detect the emission of gravitational waves (GWs) from such a binary NS progenitor. From the dynamics of the above system, the total energy emitted in GW radiation corresponds to $E^T_{GW}=7.42\times10^{52}$ erg, computed during the entire inspiral phase all the way up to the merger. This gives a signal below the sensitivity of the Advanced LIGO interferometer.

In Sec.~\ref{sec:4.4} we focus on the short-lived ($\Delta t\approx4$ s) but significant $0.1$--$100$ GeV emission (see Fig.~\ref{fig:diagram}).
We first address the issue of whether this is a peculiarity of GRB 140619B, or whether the GeV emission can be considered to be a common feature of all these \familytwo short GRBs. 
We first return to GRB 090227B to see how to explain the absence of observations of the GeV emission from this source, and we find a simple reason: GRB 090227B was outside the nominal LAT field of view \citep[FoV, see][and Sec.~\ref{GRB090227B}]{Ackermann2013}.
We then turn our attention to another source, GRB 090510, which presents many of the common features of the \familytwo short GRBs. Especially noteworthy is the presence of a high energy GeV emission lasting $\sim10^2$ s, much longer than the one of GRB 140619B.
The presence of an X-ray afterglow in GRB 090510 is  fortunate and particularly important, though lacking a scaling law behavior \citep{Ruffini2014}, since it has allowed the optical identification of the source and the determination of its distance and its cosmological redshift $z=0.903$. The corresponding isotropic energy and intrinsic peak spectral energy are, respectively, $E_{iso}>10^{52}$ erg and $E_{p,i}=(7.89\pm0.76)$ MeV, typical again of \familytwo short bursts.
We then compare and contrast this high energy emission and their corresponding X-ray emissions in the \familytwo short GRB 140619B and GRB 090510 with the afterglow of the \familyone short GRBs \citep[see Fig.~\ref{fig:8} and][]{2014ARA&A..52...43B}. 

In Sec.~\ref{sec:4.5} we give an estimate for the rate of the \familytwo short GRBs.

In Sec~\ref{sec:Calderone} we discuss the existence of the new $E_{p,i}$--$E_{iso}$ relation for all short GRBs introduced by \citet{2012ApJ...750...88Z} and \citet{Calderone2014}, with a power-law similar to the one of the Amati relation \citep{Amati2008} for long GRBs, but with a different amplitude.
Finally we draw our conclusions.

\section{Motivation from galactic binary NS and NS theory}\label{sec:NS}

Recent theoretical progress has been achieved in the understanding of the NS equation of state and equilibrium configuration and of the value of its critical mass $M_{crit}^{NS}$.
In \citet{Rotondo2011} it has been shown to be impossibile to impose the LCN condition on a self-gravitating system of degenerate neutrons, protons and electrons in $\beta$-equilibrium within the framework of relativistic quantum statistics and the Einstein-Maxwell equations. 
The equations of equilibrium of NSs, taking into account strong, weak, electromagnetic, and gravitational interactions in general relativity and the equilibrium conditions based on the Einstein-Maxwell-Thomas-Fermi equations along with the constancy of the general relativistic Fermi energies of particles, the ``Klein potentials'', throughout the configuration have been presented in \citet{Rueda2011} and \citet{Belvedere}, where a theoretical estimate of $M_{crit}^{NS}\approx 2.67~M_\odot$ has been obtained. 
The implementations of the above results by considering the equilibrium configurations of slowly rotating NSs by using the Hartle formalism has been presented in \citet{Belvedere2014}. 
Then in \citet{Rueda2014} a detailed study was made of the transition layer between the core and crust of NSs at the nuclear saturation density, and its surface tension and Coulomb energy have been calculated.
A comprehensive summary of these results for both static and uniformly-rotating NSs is discussed in \citet{BelRR}.
The absolute upper limit on the angular momentum of a rotating NS fulfilling the above microscopical conditions has been obtained in \citet{CCFRR2015}.

\begin{figure}
\centering
\includegraphics[width=\hsize,clip]{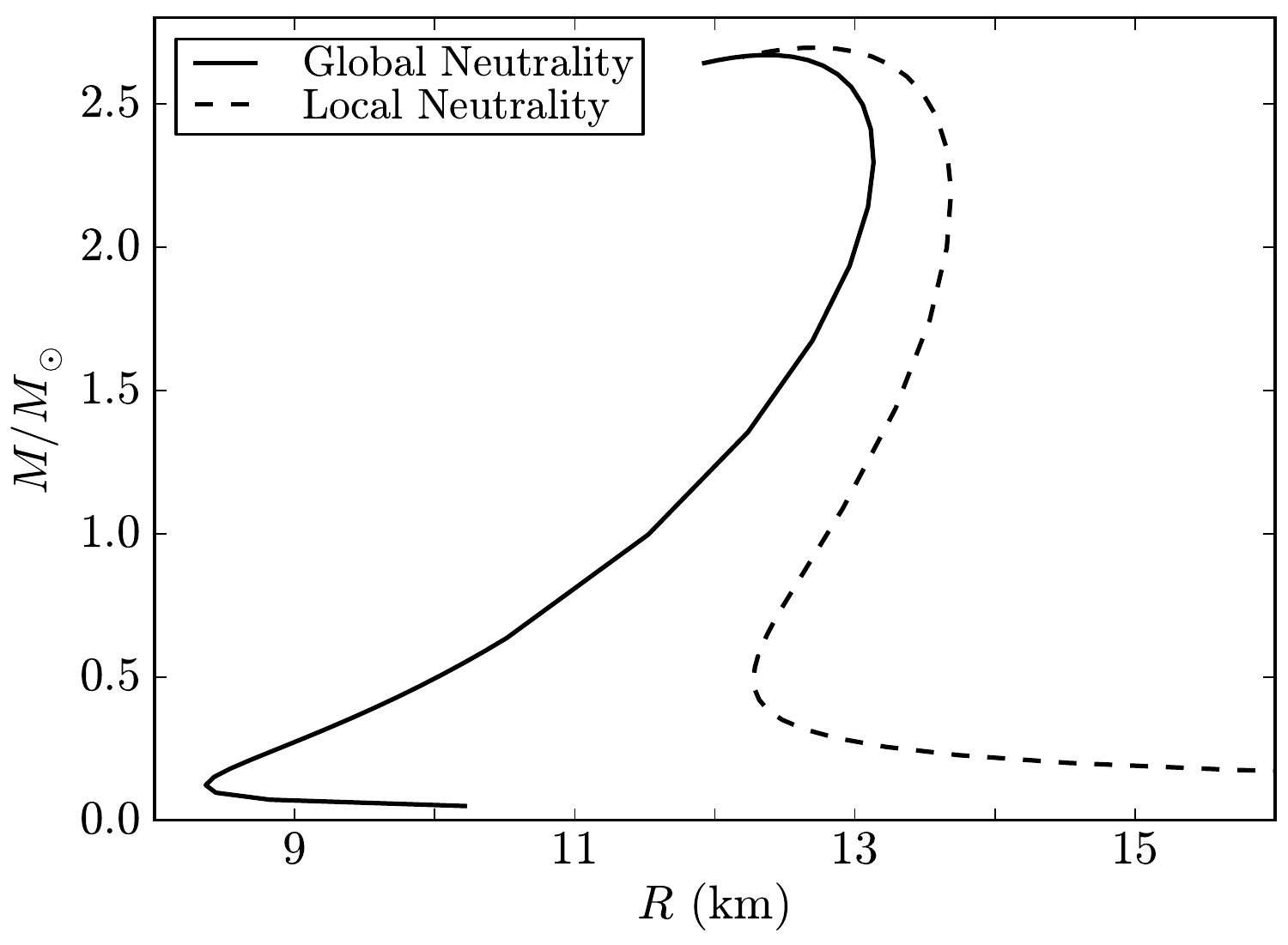}
\caption{Mass-Radius relation obtained with the local and the new global neutrality equilibrium configurations, by applying the NL3 nuclear model. Figure reproduced from \citep{Belvedere}.}
\label{fig:mass_rad}
\end{figure}

A vast number of tests have been performed in fitting the data of pulsars \citep{2012ARNPS..62..485L,2012ApJ...757...89D,2013Sci...340..448A,2014IJMPD..2330004K}.
In particular, the high value of the recently measured mass of PSR J0348+0432, $M=(2.01\pm0.04)$ M$_\odot$ \citep{2013Sci...340..448A}, favors stiff nuclear equations of state, like the one adopted in \citet{Belvedere} based on relativistic nuclear mean field theory \'a la \citet{1977NuPhA.292..413B}, which leads to the above theoretical estimate of $M_{crit}^{NS}$ (see also Fig.~\ref{fig:mass_rad}). This value is supported by the above observational constraints, and in any case, is well below the absolute upper limit of $3.2~M_\odot$ for a non-rotating NS \citep{Rhoades1974}.

\begin{figure}
\centering
\includegraphics[width=\hsize,clip]{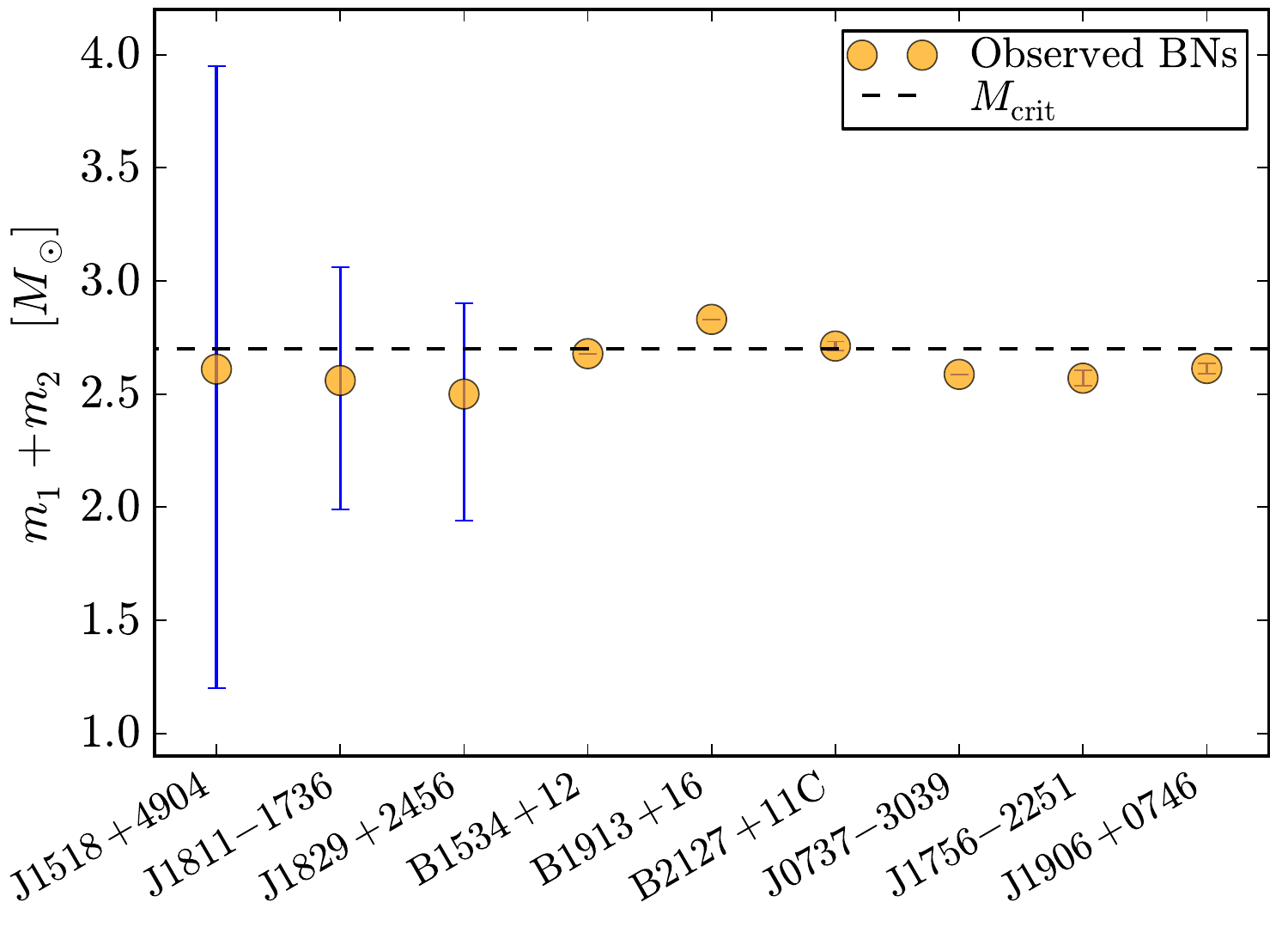}
\caption{Plot of the binary NSs with known total masses ($M_1+M_2$, in solar masses) and the corresponding uncertainties. The horizontal dashed line marks the critical NS mass of $2.67$ M$_\odot$ \citep{Belvedere}. Systems beyond this value lead to BH formation. Masses taken from \citet{2011A&A...527A..83Z} and \citet{2014arXiv1407.3404A}.}
\label{fig:dns}
\end{figure}

If we turn to the binary NSs within our Galaxy (see Fig.~\ref{fig:dns}) we notice that only in a subset of them is the total mass of the components larger than $M_{crit}^{NS}$ and can lead to a BH in their merging process%
\footnote{During the refereeing process, an approach by \citet{2015arXiv150407605F} based on a combination of binary NS merger nuclear physics models and population synthesis appeared. They infer that for a maximum nonrotating NS mass of $M_{crit}^{NS}$ above $2.3$--$2.4$ M$_\odot$, less than $4\%$ of the NS mergers produces short GRBs by gravitational collapse to a BH. 
Here we go one step further by indicating the theoretical predictions characterizing short GRBs originating from the massive NS formation (family-1) and the ones originating from BH formation (family-2). We indicate: a) the specific spectral features, b) the presence of the GeV emission originating from the BH, and c) the fulfillment of the $E_{p,i}$--$E_{iso}$ relation \citep[see][and Sec.~\ref{sec:Calderone}]{2012ApJ...750...88Z,Calderone2014}.
The paper by \citet{2015arXiv150407605F} was followed by \citet{2015arXiv150500231L} where the authors examine the value of $M_{crit}^{NS}$ for a family of equations of state and concluded that a reasonable fraction of double NS mergers may produce neither short GRBs nor BHs. Here we again go one step further by indicating that in the case of a merged core with a mass smaller than $M_{crit}^{NS}$ leading to a massive NS, a less energetic short GRB with a softer emission tail indeed occurs (\familyone short bursts). We show also that these short GRBs fulfill the above $E_{p,i}$--$E_{iso}$ relation (see Sec.~\ref{sec:Calderone}).}.

Given this general understanding, we have identified the characteristic properties of \familytwo short bursts, whose prototype was identified in GRB 090227B \citep{Muccino2013}. Equally important has been the identification of the observed characteristic features of \familyone short GRBs which will be discussed in the following sections.

The crucial role of $M_{crit}^{NS}$ has been also shown in the corresponding analysis of long GRBs in distinguishing between the two different families \citep{Yu2014} in the induced gravitational collapse (IGC) paradigm \citep{IGC,IGC2,Fryer2014}.

\section{The fireshell model}\label{sec:2}

It is well known that the majority of the astrophysical community working on GRBs envisages the spectral and temporal analysis of both short and long GRBs considering their whole emission as a single event \citep[see, e.g.,][]{Ackermann2013}. This picture follows the conceptual framework of the ``fireball model'' \citep[see, e.g.,][and reference therein]{Sari1998b,Piran2005,Meszaros2006}.

The ``fireshell model" \citep{Ruffini2001c,Ruffini2001,Ruffini2001a} has instead addressed a specific time-resolved spectral analysis leading to distinct signatures and to the identification of different astrophysical regimes within the same GRB \citep[see, e.g.,][and references therein]{Izzo2010,Izzo2012,Muccino2013,Rees_nuovo}.
This has led to introduction of the concept of binary mergers of NS--NS and of FeCO--NS together with a set of new paradigms in order to describe the complexity of GRB phenomena within a ``Cosmic-Matrix" approach \citep{RuffiniZeldovich}.

In the fireshell model \citep{Ruffini2001c,Ruffini2001,Ruffini2001a} GRBs originate from an optically thick $e^+e^-$ plasma 
\citep{Damour,Xue2008,RVX2010} during the gravitational collapse to a BH. 
Such an $e^+e^-$ plasma is confined to an expanding shell and reaches thermal equilibrium almost instantaneously \citep{2007PhRvL..99l5003A}. The annihilation of these pairs occurs gradually, while the expanding shell, called the \textit{fireshell}, self-accelerates up to ultra relativistic velocities \citep{RSWX2} and engulfs the baryonic matter (of mass $M_B$) left over in the process of collapse. 
The baryon load thermalizes with the pairs due to the large optical depth \citep{RSWX}. 

\begin{figure}
\centering
\includegraphics[width=\hsize,clip]{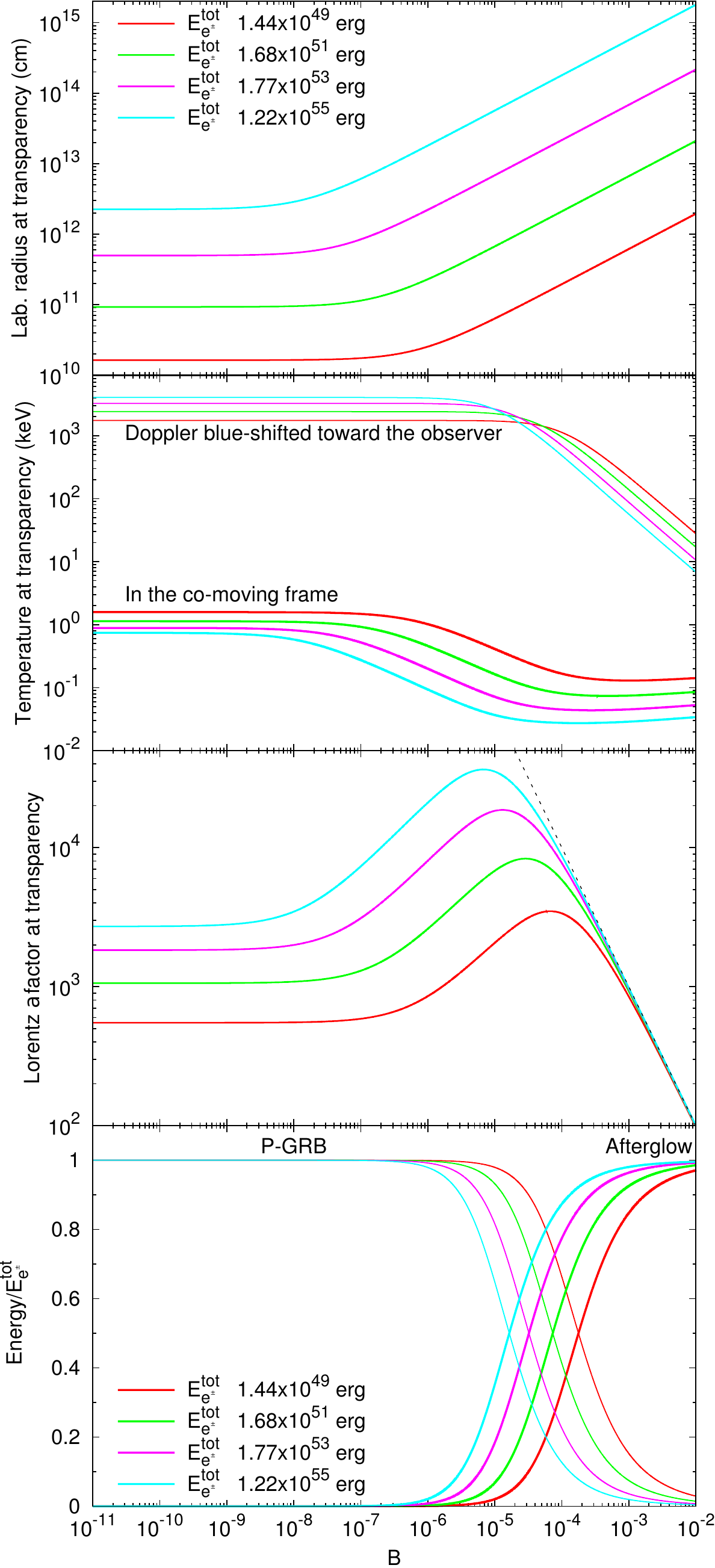}
\caption{The main quantities of the fireshell model at transparency for selected values of $E_{e^+e^-}^{tot}$: the radius in the laboratory frame, the temperatures of the plasma in the co-moving frame and blue-shifted toward the observer, the Lorentz $\Gamma$ factor, and the fraction of energy radiated in the P-GRB and in the prompt emission as functions of $B$.}
\label{fig:0}
\end{figure}

Assuming spherical symmetry of the system, the dynamics in the optically thick phase is fully described by only two free initial parameters: the total energy of the plasma $E^{tot}_{e^+e^-}$ and the baryon load $B$ \citep{RSWX}.
Only solutions with $B\leq10^{-2}$ are characterized by regular relativistic expansion; for $B\geq10^{-2}$ turbulence and instabilities occur \citep{RSWX}.
The fireshell continues to self-accelerate until it reaches the transparency condition and a first flash of thermal radiation, the P-GRB, is emitted \citep{Ruffini2001}. 
The radius $r_{tr}$ at which the transparency occurs, the theoretical temperature (blue-shifted toward the observer $kT_{blue}$), the Lorentz factor $\Gamma_{tr}$, as well as the amount of the energy emitted in the P-GRB are functions of $E_{e^+e^-}^{tot}$ and $B$ \citep[see, e.g.,][and  Fig.~\ref{fig:0}]{Ruffini2001,Ruffini2009b}.

After transparency, the residual expanding plasma of leptons and baryons collides with the CBM giving rise to multi-wavelength emission: the prompt emission.
Assuming the fully-radiative condition, the structures observed in the prompt emission of a GRB are described by two quantities associated with the environment: the CBM density profile $n_{CBM}$, which determines the temporal behavior of the light curve, and the fireshell surface filling factor $\mathcal{R}=A_{eff}/A_{vis}$, in which $A_{eff}$ is the effective emitting area of the fireshell, and $A_{vis}$ is its total visible area \citep{Ruffini2002,Ruffini2005}. 
This second parameter takes into account the inhomogeneities in the CBM and its filamentary structure \citep{Ruffini2004}.

The emission process of the collision between the baryons and the CBM is described in the comoving frame of the shell as a modified black body (BB) spectrum. 
This spectrum is obtained by the introduction of an additional phenomenological parameter $\alpha$ which characterizes the departure of the slope of the low energy part of the comoving spectrum from the purely thermal one \citep[see][for details]{Patricelli}.
The nonthermal spectral shape of the observed GRB is then produced by the convolution of a very large number of modified thermal spectra with different temperatures and different Lorentz and Doppler factors.
This convolution is performed over the surfaces of constant arrival time for photons at the detector \cite[EQuiTemporal Surfaces, EQTS,][]{Bianco2005b,Bianco2005a}, encompassing the total observation time. 
The observed hard-to-soft spectral variation comes out naturally from the decrease with time of the comoving temperature and of the bulk Lorentz $\Gamma$ factor. 
This effect is amplified by the curvature effect due to the EQTS which produces the observed time lag in the majority of the GRBs.

The canonical GRB light curve within the fireshell model is then characterized by a first (mainly thermal) emission due to the transparency of the $e^+e^-$-photon-baryon plasma, the P-GRB.
A multi-wavelength emission, the prompt emission, follows due to the collisions between the accelerated baryons and the CBM.

The fireshell model has originally described the process of vacuum polarization due to the overcritical electromagnetic field occurring at the moment of BH formation \citep{Damour}. The formalism has been developed by considering a large number of relativistic quantum effects in the electrodynamics proposed for the NS crust \citep{Belvedere,Belvedere2014,RuedaRRWu2014}, as well as on quantum-electrodynamics processes ongoing in the gravitational collapse \citep{2012PhRvD..86h4004H,2013PhLA..377.2450R}.
This has led to the results summarized in Fig.~\ref{fig:0}.

The first description of the $e^+e^-$ plasma within the fireshell model was performed under the simplified assumption of spherical symmetry \citep[the dyadosphere, see, e.g.,][]{Preparata}. The corresponding structure in the axially symmetric Kerr-Newman geometry has been considered  \citep[the dyadotorus, see, e.g.,][]{Cherubini,RRKerr} and could possibly be tested.

The general formalism of the fireshell model can also be applied to any optically thick $e^+e^-$ plasma in the presence of a baryon load, like the one created during the merging of binary NSs from $\nu\bar{\nu}\to e^+e^-$ \citep[see, e.g.,][and references therein]{SalmonsonWilson2002}.

The  P-GRB addresses the fully relativistic fundamental physics aspects of the model, in particular the acceleration process of the $e^+e^-$-baryon plasma, the collapsing NS quantum-electrodynamics, and the BH physics. The prompt emission addresses the conceptually simpler problem of the interaction of the accelerated baryons with the CBM, which does not allow nor require, by its own nature, a detailed description.

\section{Summary of the results for GRB 090227B: the prototype of the family-2 short GRBs}\label{GRB090227B}

GRB 090227B is a bright short burst with an overall emission lasting $\sim 0.9$ s and total fluence of $3.79 \times 10^{-5}$ erg/cm$^2$ in the energy range $8$ keV -- $40$ MeV.
This burst was significantly detected only in the LAT Low Energy (LLE)  data since it was outside the nominal LAT field of view (FoV) \citep{Ackermann2013}. However, only one transient-class event with energy above $100$ MeV has been associated with the GRB \citep{Ackermann2013}.

The time-resolved spectral analysis on the time scale as short as $16$ ms, made possible by the \textit{Fermi}--GBM \citep{Meegan2009}, has allowed the identification of the P-GRB in the early $96$ ms of emission.
The corresponding thermal component has a temperature $kT=(517\pm28)$ keV (see the upper plots of Fig.~9 in \citealp{Muccino2013}). 
The subsequent emission, fit by a Band function (see lower plots of Fig.~9 in \citealp{Muccino2013}), has been identified with the prompt emission.

Due to the absence of an optical identification, a direct measurement of the cosmological redshift was not possible.
From the temperature and flux of the P-GRB thermal component it was possible to derive (see Fig.~\ref{fig:0}) a theoretical cosmological redshift $z=1.61\pm0.14$, as well as the baryon load $B=(4.13\pm0.05)\times10^{-5}$, the total plasma energy $E^{tot}_{e^+e^-}=(2.83\pm0.15)\times10^{53}$ erg, and the extremely high Lorentz $\Gamma$ factor at transparency $\Gamma_{tr}=(1.44\pm0.01)\times10^4$ (see Sec.~4.1 in \citealp{Muccino2013}). 
Consequently, an average CBM number density $\langle n_{CBM}\rangle = (1.90\pm0.20)\times10^{-5}$ cm$^{-3}$ has been determined which is typical of galactic halos where NS--NS mergers migrate, owing to natal kicks imparted to the binaries at birth \citep[see, e.g.,][]{2014ARA&A..52...43B,Narayan1992,1999MNRAS.305..763B,1999ApJ...526..152F,2006ApJ...648.1110B}.

In \citet{Muccino2013} it was concluded that the progenitor of GRB 090227B is a binary NS. For simplicity and as a lower limit, the masses of the two NS have been assumed to be the same, e.g. $M_1=M_2=1.34$ M$_\odot$, so that the total merged core mass is $>M_{crit}^{NS}$ and therefore a BH is formed. This conclusion was drawn in view of the large total energy, $E^{tot}_{e^+e^-}=2.83\times10^{53}$ erg.
Correspondingly, the energy emitted via gravitational waves, $\sim 9.7\times10^{52}$ erg, has been estimated in \citet{Oliveira2014}.

\section{Observations and data analysis of GRB 140619B}\label{sec:3}

At 11:24:40.52 UT on $19^{th}$ June 2014, the \textit{Fermi}-GBM detector \citep{GCN16419} triggered and located the short and hard burst GRB 140619B (trigger 424869883/140619475). 
The on-ground calculated location, using the GBM trigger data, was RA(J$2000$)$=08^h54^m$ and Dec(J$2000$)$=-3^{\rm{o}}42^\prime$, with an uncertainty of $5^{\rm{o}}$ (statistical only).  
The location of this burst was 32$^{\rm{o}}$ from the LAT boresight at the time of the trigger, and the data from the \textit{Fermi}-LAT showed a significant increase in the event rate \citep{GCN16421}.
The burst was also detected by \textit{Suzaku}-WAM \citep{GCN16457}, showing a single pulse with a duration of $\sim 0.7$ s ($50$ keV -- $5$ MeV).
The analysis from $48.7$ ks to $71.6$ ks after the GBM trigger by the \textit{Swift}-XRT instrument in the field of view of the \textit{Fermi}-GBM and LAT, was completely in Photon Counting (PC) mode \citep{GCN16424}. No bright X-ray afterglow was detected within the LAT error circle. This set an upper limit on the energy flux in the observed $0.3$--$10$ keV energy band of $\approx9.24\times10^{-14}$ erg/(cm$^2$s), assuming a photon index $\gamma=2.2$. Therefore, no optical follow-up was possible and thus the redshift of the source is unknown. 

\begin{figure}
\centering
\includegraphics[width=\hsize,clip]{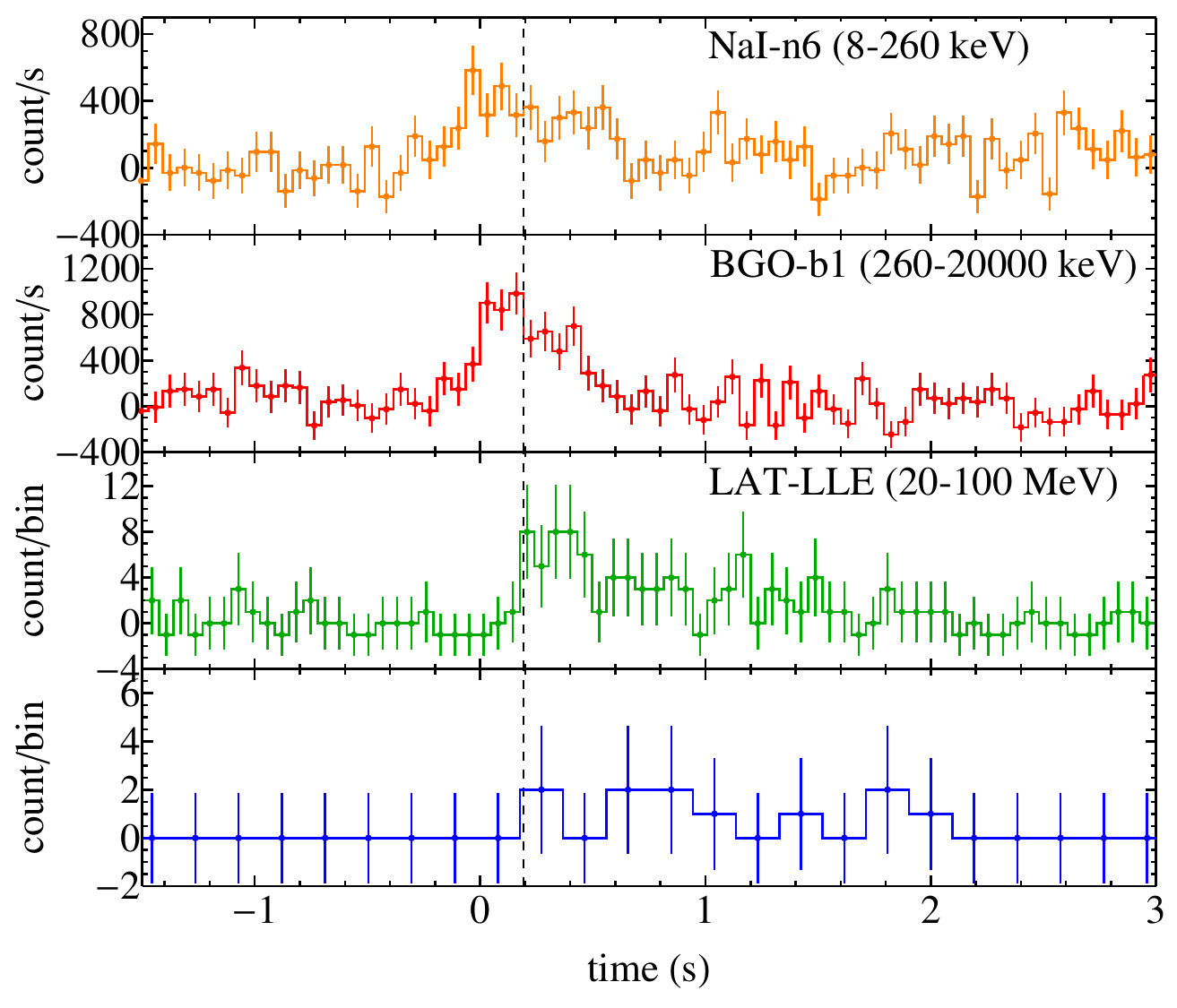}
\caption{Background subtracted light curves of GRB 140619B from various detectors in various energy bands. From the top to the bottom panel: the $64$ ms binned light curves from the NaI-n6 ($8$ -- $260$ keV, top panel) and BGO-b1 ($260$ keV -- $20$ MeV, second panel) detectors, the $64$ ms binned LLE light curve ($20$ -- $100$ MeV, third panel) and the $192$ ms binned high-energy channel light curve ($100$ MeV -- $100$ GeV, bottom panel).}
\label{fig:1}
\end{figure}

We have analyzed the \textit{Fermi}-GBM and LAT data in the energy range $8$ keV -- $40$ MeV and $20$ MeV -- $100$ GeV, respectively. 
We have downloaded the GBM \texttt{TTE} (Time-Tagged Events) files%
,\footnote{ftp://legacy.gsfc.nasa.gov/fermi/data/gbm/bursts} 
suitable for short or highly structured events, and analyzed them by using the \texttt{RMFIT} package\footnote{http://\textit{Fermi}.gsfc.nasa.gov/ssc/data/analysis/rmfit/vc$\_$rmfit$\_$tutorial.pdf}.
The LAT Low Energy (LLE) data%
\footnote{http://fermi.gsfc.nasa.gov/ssc/observations/types/grbs/lat$\_$grbs/}, between $20$ -- $100$ MeV, and the high energy data\footnote{http://fermi.gsfc.nasa.gov/cgi-bin/ssc/LAT/LATDataQuery.cgi}, between $100$ MeV -- $100$ GeV, were analyzed by using the Fermi-Science tools.
\footnote{http://fermi.gsfc.nasa.gov/ssc/data/analysis/documentation/Cicerone/}
In Fig.~\ref{fig:1} we have reproduced the $64$ ms binned GBM light curves corresponding to detectors NaI-n6 ($8$ -- $260$ keV, top panel) and BGO-b1 ($260$ keV -- $20$ MeV, second panel), the $64$ ms binned LLE light curve ($20$ -- $100$ MeV, third panel) and the $192$ ms binned high-energy channel light curve ($0.1$ -- $100$ GeV, bottom panel). 
All the light curves are background subtracted.
The NaI-n6 light curve shows a very weak signal, almost at the background level, while the BGO-b1 signal is represented by a short hard pulse, possibly composed by two sub-structures, with a total duration of $T_{90}\approx0.7$ s. The vertical dashed line in Fig.~\ref{fig:1} represents the on-set of both LAT light curves, i.e., $\sim0.2$ s after the GBM trigger.
In principle, this allows us to determine the time interval within which the P-GRB emission takes place.

We have subsequently performed the time-integrated and time-resolved spectral analyses focused on the GBM data in the energy range $8$ keV -- $40$ MeV.

\subsection{Time-integrated spectral analysis}\label{sec:3.1}

\begin{figure*}
\centering 
\hfill
\includegraphics[width=0.49\hsize,clip]{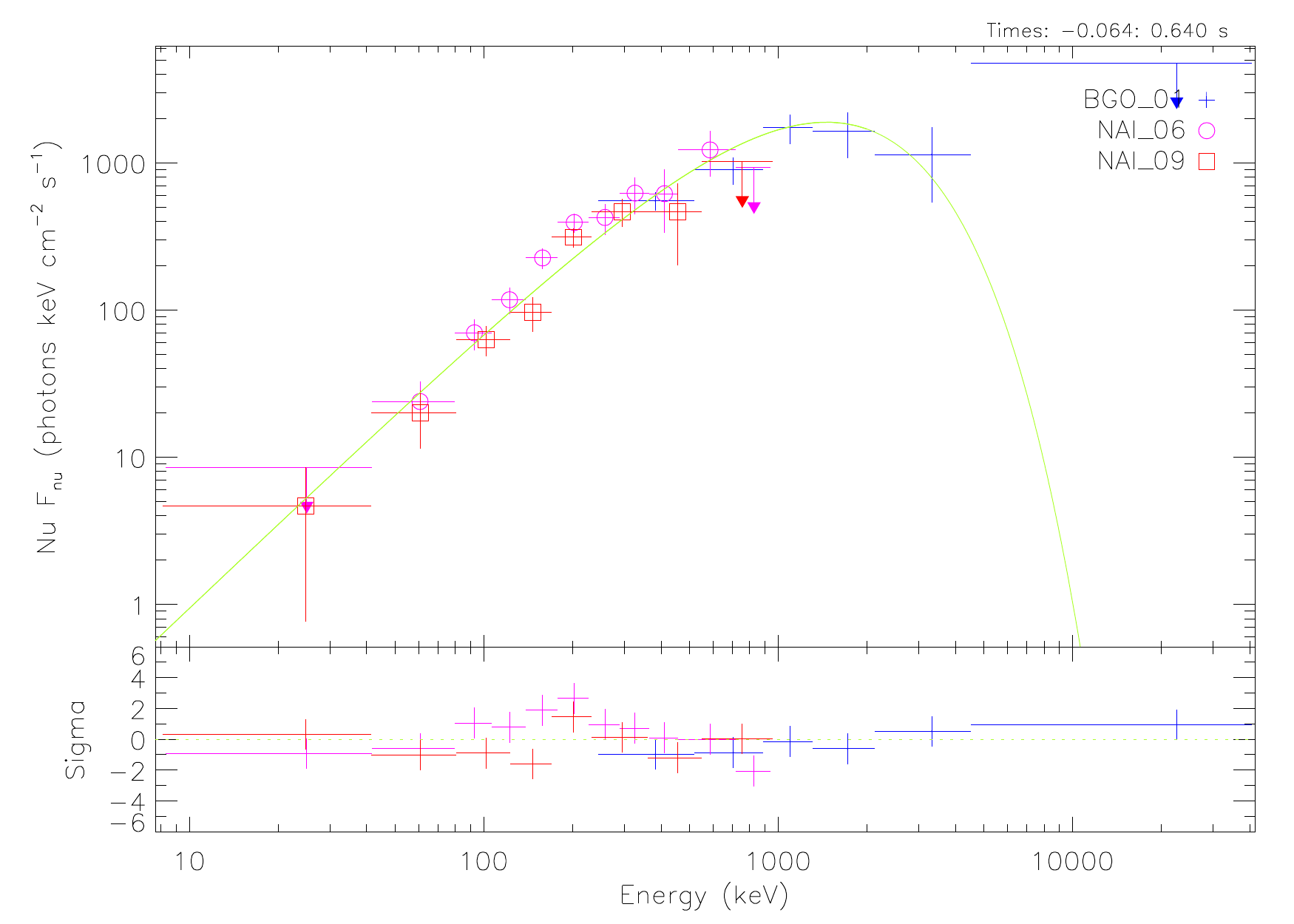}
\includegraphics[width=0.49\hsize,clip]{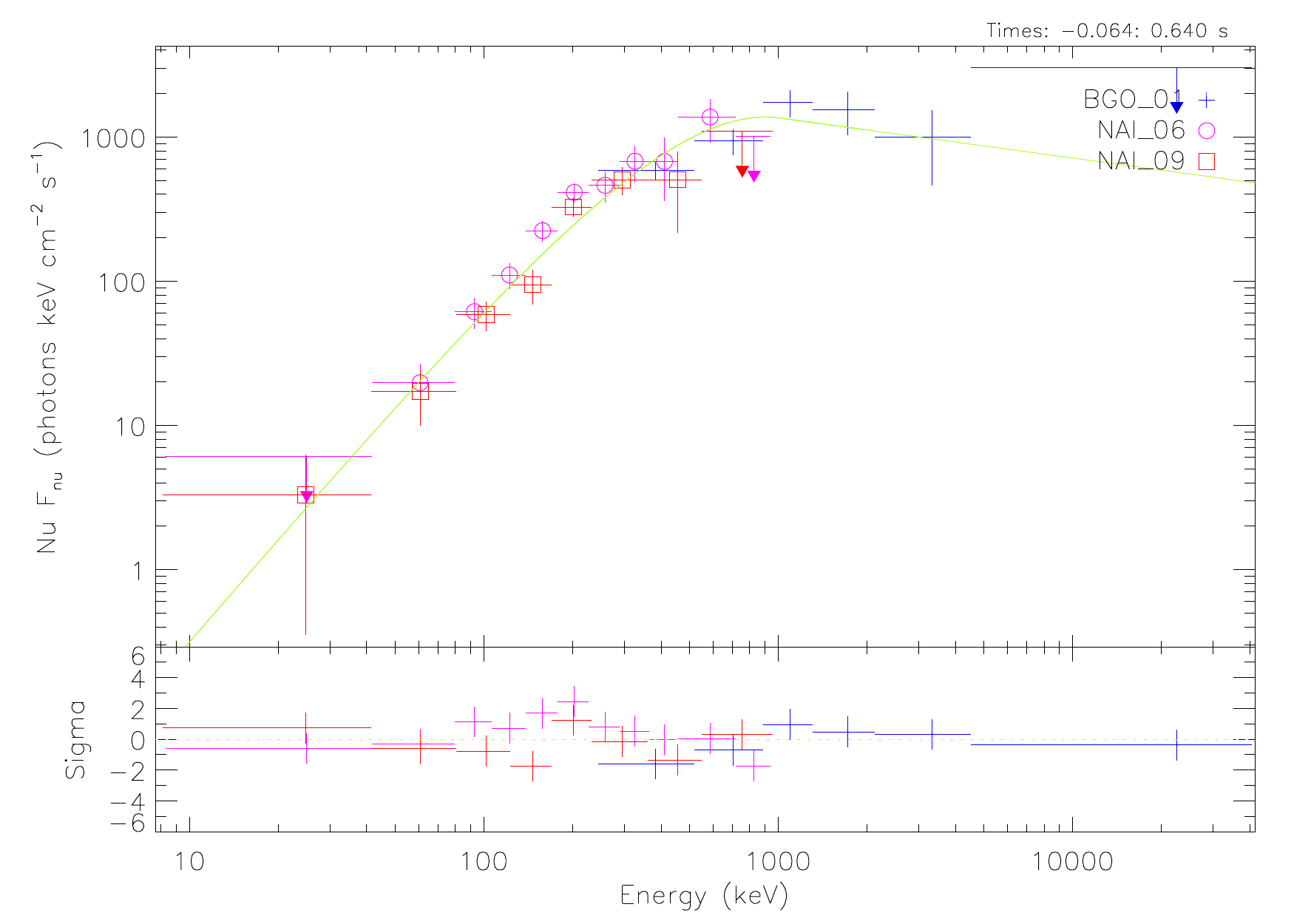} 
\caption{The combined NaI--n6, n9+BGO--b1 $\nu F_\nu$ spectra of GRB 140619B in the $T_{90}$ time interval. The fit using the Compt spectral model is shown on the left, while the Band model fit is on the right.}
\label{fig:2}
\end{figure*}

We have performed a time-integrated spectral analysis in the time interval from $T_0-0.064$ s to $T_0+0.640$ s, which corresponds to the $T_{90}$ duration of the burst. We have indicated  the trigger time by $T_0$ and have considered the following spectral models: Comptonization (Compt) and a Band function \citep{Band1993}.
The corresponding plots are shown in Fig.~\ref{fig:2} and the results of the fits are listed in Tab.~\ref{tab:1}.
From a statistical point of view, the Compt model provides the best fit to the data. In fact the Band function, which has an additional parameter with respect to the Compt model, improves the fit by only $\Delta$C-STAT$=2.53$, where $\Delta$C-STAT is the difference between the two C-STAT values of the Compt and Band models. If we consider $\Delta$C-STAT as a $\chi^2$ variable for the change in the number of the model parameters $\Delta n$ (in this case $\Delta n=1$), and assuming that the Compt model is nested within the Band model%
,\footnote{The Compt model can be considered a particular case of the Band model with $\beta\rightarrow-\infty$.} 
we conclude that the Band model improves the fit only at the $89\%$ significance level, and anyway less than 2 $\sigma$.
Therefore it is not enough to reject the Compt model.
The most interesting feature of the Compt model consists of its low-energy index, which is consistent with $\alpha\sim0$. 
We proceed now to a time-resolved analysis to investigate the possibility that in the early phases of the prompt emission the spectrum is consistent with a BB spectrum, i.e., $\alpha\approx1$, which corresponds to the signature of P-GRB emission.
\begin{table*}
\scriptsize
\centering
\begin{tabular}{ccccccccc}
\hline\hline
$\Delta T$      &  Model  &  $K$ (ph keV$^{-1}$ cm$^{-2}$s$^{-1}$)  &  $kT$ (keV)    &  $E_{p}$ (keV)   &  $\alpha$        & $\beta$                 &  $F_{tot}$ (erg cm$^{-2}$s$^{-1}$)      &  C-STAT/DOF    \\
\hline
$T_{90}$        &  Compt  &  $(7.7\pm1.1)\times10^{-3}$             &                &  $1456\pm216$    &  $-0.09\pm0.18$  &                         &  $(5.75\pm0.75)\times 10^{-6}$          &  $365.09/346$  \\
                &  Band   &  $(7.8\pm1.3)\times10^{-3}$             &                &  $908\pm199$     &  $-0.38\pm0.37$  & $-2.28\pm0.31$          &  $(7.4\pm1.8)\times 10^{-6}$            &  $362.56/345$  \\
\hline
$\Delta T_{1}$  &  Compt  &  $(6.3\pm2.0)\times10^{-3}$             &  &  $1601\pm287$  &  $0.26\pm0.32$   &                                       &  $(9.4\pm1.6)\times 10^{-6}$            &  $318.92/346$  \\
                &  BB     &  $(7.5\pm2.2)\times10^{-8}$             &  $324\pm33$    &                   &              &                         &  $(8.5\pm1.2)\times 10^{-6}$            &  $323.86/347$  \\
\hline
$\Delta T_{2}$  &  Compt  &  $(7.2\pm1.4)\times10^{-3}$           &    &  $1283\pm297$  &  $-0.11\pm0.26$  &                                   &  $(4.38\pm0.89)\times 10^{-6}$          &  $391.65/346$  \\
                &  BB     &  $(3.8\pm1.1)\times10^{-7}$               &  $156\pm15$    &                            &         &                &  $(2.33\pm0.28)\times 10^{-6}$          &  $392.23/347$  \\
\hline
\end{tabular}
\caption{Summary of the time-integrated ($T_{90}$) and time-resolved ($\Delta T_1$ and $\Delta T_2$) spectral analyses. In each column are listed, respectively, the time interval $\Delta T$, the adopted spectral model, the normalization constant $K$ of the fitting function, the BB temperature $kT$, the peak energy $E_p$, the low-energy $\alpha$ and high-energy $\beta$ photon indexes, the total energy flux $F_{tot}$ in the range $8$ keV -- $40$ MeV, and the value of the C-STAT over the number of degrees of freedom (DOF).}
\label{tab:1}
\end{table*}

\subsection{Time-resolved spectral analysis}\label{sec:3.2}

We performed the time-resolved spectral analysis by selecting time intervals with fluences larger than $\approx10^{-6}$ erg cm$^{-2}$ in order to collect enough photons. Consequently, we have selected two time intervals that correspond to the main spike and the less intense structure (see the BGO-b1 light curve in Fig.~\ref{fig:2}). The first time interval, from $T_0$ to $T_0+0.192$ s, is hereafter referred to as $\Delta T_1$, while the subsequent emission, from $T_0+0.192$ s to $T_0+0.640$ s, is designated by $\Delta T_2$.

\begin{figure*}
\centering 
\hfill
\includegraphics[width=0.49\hsize,clip]{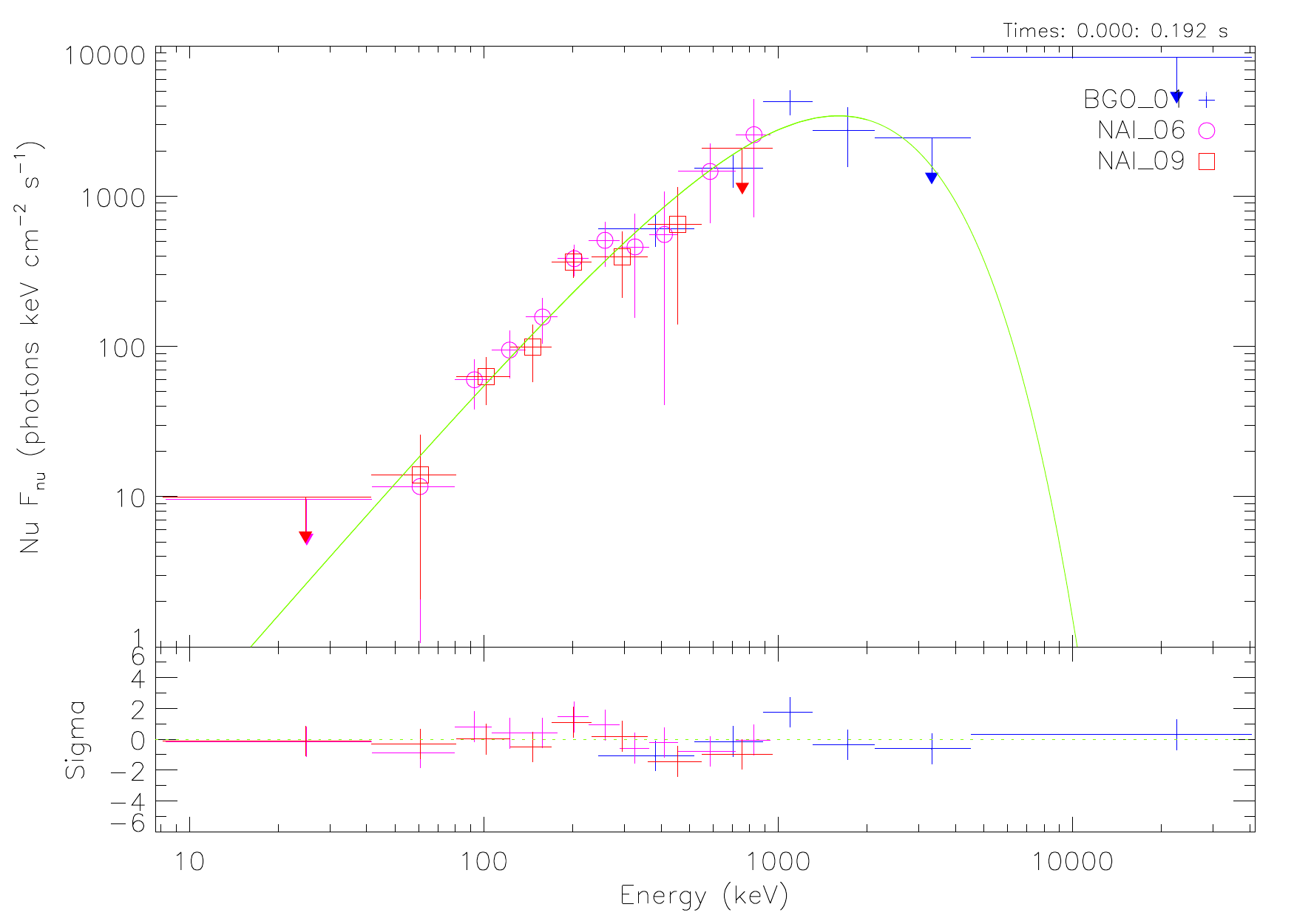}
\includegraphics[width=0.49\hsize,clip]{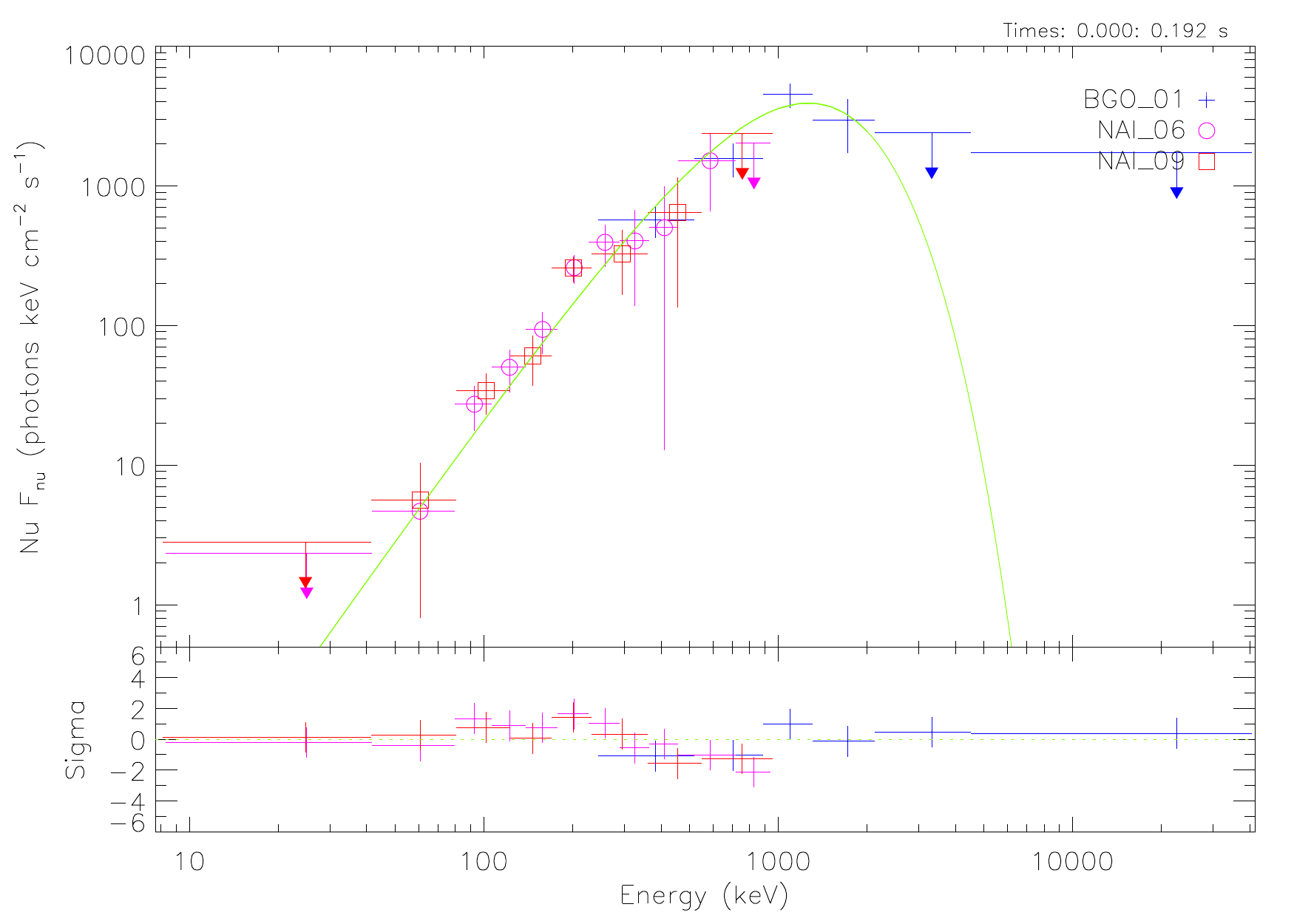} 
\caption{The same considerations as in Fig.~\ref{fig:2}, in the $\Delta T_1$ time interval, comparing Compt (left panel) and BB (right panel) models.}
\label{fig:3}
\end{figure*}

In the $\Delta T_1$ time interval, to identify the P-GRB, we have performed a spectral analysis  by considering the BB and Compt spectral models. The spectra and the corresponding fits are shown in Fig.~\ref{fig:3} and the best fit parameters are listed in Tab.~\ref{tab:1}.
As reported in Tab.~\ref{tab:1}, the Compt and the BB models are both viable. 
However, the value of the low-energy index of the Compt model in the $\Delta T_1$ time interval, $\alpha=0.26\pm0.32$, is consistent within three $\sigma$ with $\alpha=1$, which is the low energy index of a BB. 
We conclude that the BB model is an acceptable fit to the data and the best ``physical model'' of the $\Delta T_1$ time interval and therefore identify it with the P-GRB emission.
The corresponding observed temperature is $kT=(324\pm33)$ keV (see Tab.~\ref{tab:1}).

\begin{figure*}
\centering 
\hfill
\includegraphics[width=0.49\hsize,clip]{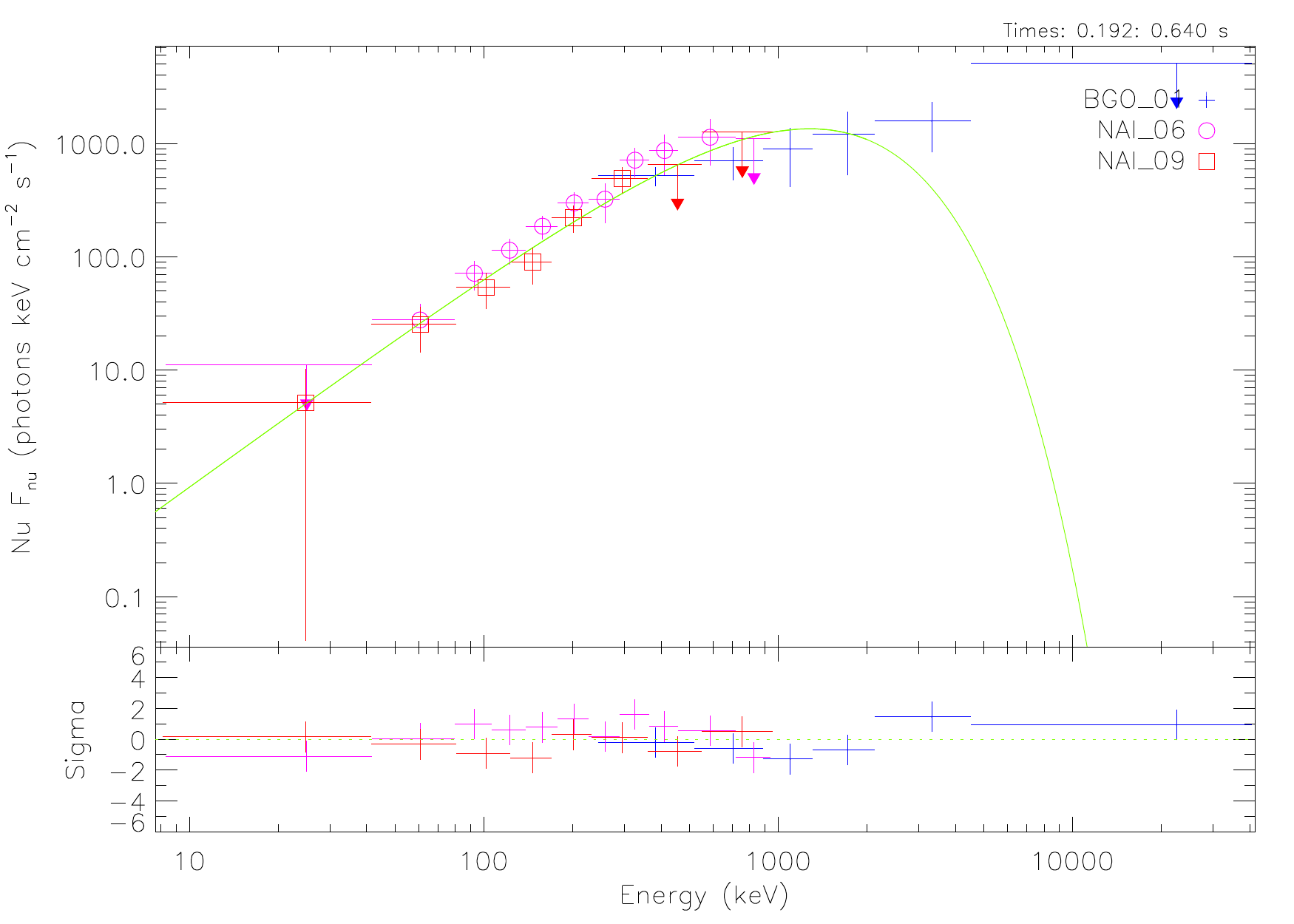}
\includegraphics[width=0.49\hsize,clip]{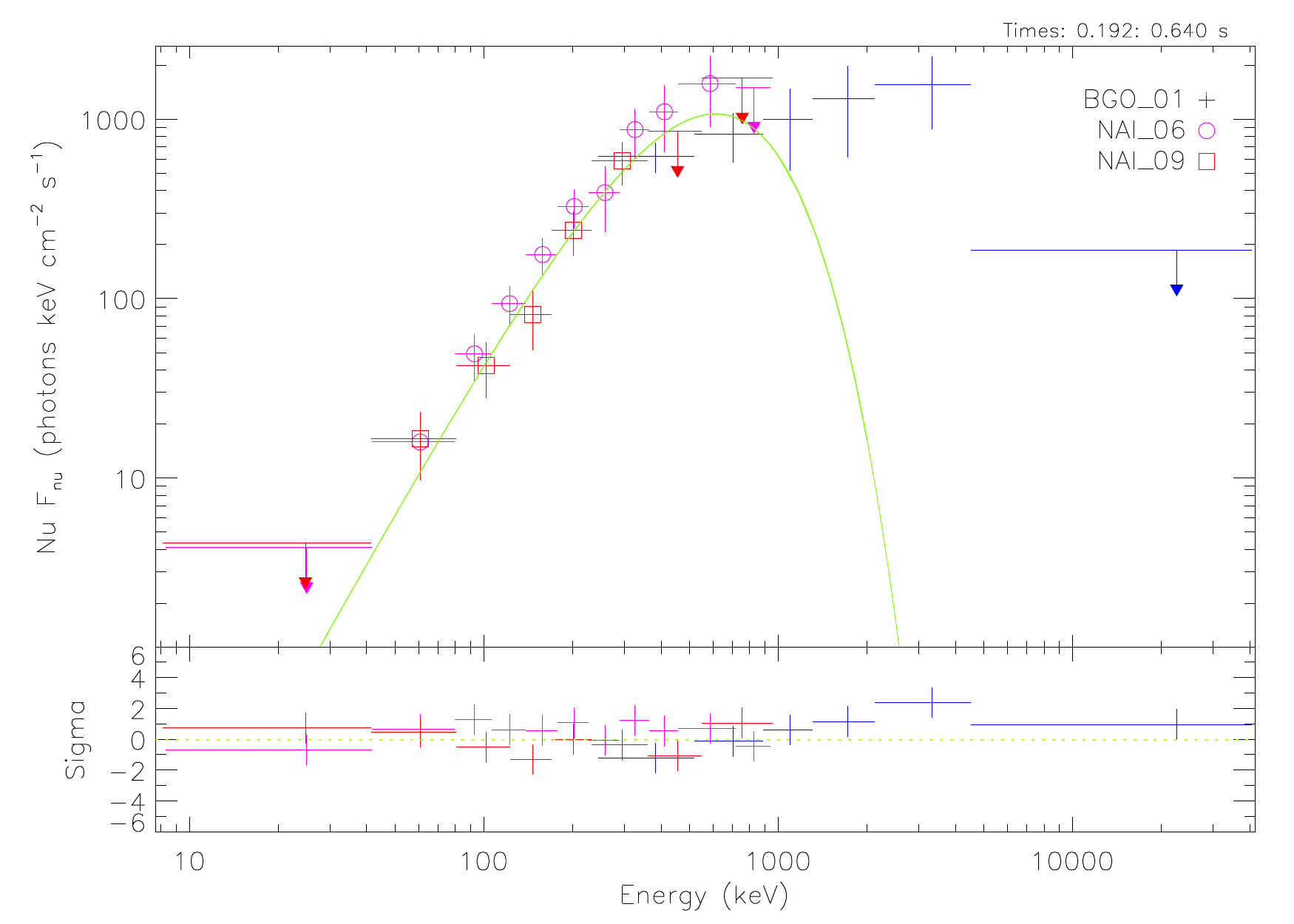} 
\caption{The same considerations as in Fig.~\ref{fig:2}, in the $\Delta T_2$ time interval, comparing Compt (left panel) and BB (right panel) models.}
\label{fig:4}
\end{figure*}

We then performed a spectral analysis on the time interval $\Delta T_2$ to identify the prompt emission. 
We have again considered the Compt and BB spectral models (see Fig.~\ref{fig:4} and Tab.~\ref{tab:1}). 
By looking at  Fig.~\ref{fig:4}, it is immediately clear that the BB model does not adequately fit the data at energies larger than $1$ MeV.
Therefore the Compt model is favored. Its low-energy index, $\alpha=-0.11\pm0.26$, indicates that the spectral energy distribution in the $\Delta T_2$ time interval is broader than that of the BB model.
The Compt model is consistent with the spectral model adopted in the fireshell model and described in \citet{Patricelli} for the prompt emission.

In the next Section we interpret the above data within the fireshell theoretical framework.

\section{Application of the fireshell model to GRB 140619B}\label{sec:4}

After the P-GRB and the prompt emission identification, we have followed the same analysis described in \citet{Muccino2013} to determine the cosmological redshift, the baryon load and all the other physical quantities characterizing the plasma at the transparency point (see Fig.~\ref{fig:0}).
It is appropriate to underline that a remarkable difference between the long and the short GRBs is considered: the P-GRB emission in long GRBs represents on average the $1$--$5\%$ of the overall emission (see, e.g., the cases of GRB 970828, \citealp{Rees_nuovo}, and GRB 090618, \citealp{Izzo2012}), while in the cases of the short GRBs 090227B and 140619B (see Sec.~\ref{sec:4.1}), the P-GRB emissions represent $\sim40\%$ of the overall observed fluence.

\subsection{Redshift estimate in fireshell model}\label{sec:4.1}

From the observed P-GRB and total fluences, respectively, $S_{BB}=F_{tot}(\Delta T_1)\Delta T_1$ and $S_{tot}=F_{tot}(T_{90})T_{90}$ (see values in Tab.~\ref{tab:1}), we have estimated the ratio
\begin{equation}
\label{ratio}
\frac{E_{P-GRB}}{E_{e^+e^-}^{tot}} \approx \frac{4\pi d_l^2 S_{BB}/(1+z)}{4\pi d_l^2 S_{tot}/(1+z)} = \frac{S_{BB}}{S_{tot}} = (40.4\pm7.8)\%\ ,
\end{equation}
where the theoretically-computed energy of the P-GRB, $E_{P-GRB}$, has been constrained by the observed thermal emission, $E_{BB}$, and we have imposed $E_{e^+e^-}^{tot} \equiv E_{iso}$. 
In Eq.~\ref{ratio} the luminosity distance $d_l$ and the redshift $z$ of the source do not enter into the final computation.

From the last diagram in Fig.~\ref{fig:0}, it is clear that for the value in Eq.~(\ref{ratio}), we have different possible parameters ($E^{tot}_{e^+e^-}$, $B$) and for each of them we can determine the corresponding $kT_{blue}$ (see the top diagram in Fig.~\ref{fig:0}). Finally, from the ratio between $kT_{blue}$ and the observed P-GRB temperature $kT$, we can estimate the redshift, i.e., $kT_{blue}/kT=(1+z)$.
To obtain the correct value of $z$ and then the right parameters $[E_{e^+e^-}^{tot}(z),B(z)]$, we have made use of the isotropic energy formula
\begin{equation}
\label{correction}
E_{iso} = 4\pi d_l^2 \frac{S_{tot}}{(1+z)} \frac{\int^{10000/(1+z)}_{1/(1+z)}{E\,N(E) dE}}{\int^{40000}_{8}{E\,N(E) dE}}\ ,
\end{equation}
in which $N(E)$ is the photon spectrum of the burst and the integrals are due to the $K$-correction on $S_{tot}$ \citep{Schaefer2007}.
From the initial constraint $E_{iso} \equiv E_{e^+e^-}^{tot}$, we have found $z =2.67\pm0.37$, which leads to $B = (5.52\pm0.73)\times10^{-5}$ and $E_{e^+e^-}^{tot} = (6.03\pm0.79)\times10^{52}$ ergs.
All the quantities so determined are summarized in Tab.~\ref{tab:2}.
The analogy with the prototypical source GRB 090227B, for which we have $E_{P-GRB}=(40.67\pm0.12)\%E_{e^+e^-}^{tot}$ and $B=(4.13\pm0.05)\times10^{-5}$, is very striking \citep{Muccino2013}.
\begin{table}
\centering
\begin{tabular}{cc}
\hline\hline
Fireshell Parameter                    &  Value                          \\
\hline
$E^{tot}_{e^+e^-}$\,(erg)              &  $(6.03\pm0.79)\times10^{52}$   \\
$B$                                    &  $(5.52\pm0.73)\times10^{-5}$   \\
$\Gamma_{tr}$                          &  $(1.08\pm0.08)\times10^4$      \\
$r_{tr}$\,(cm)                         &  $(9.36\pm0.42)\times10^{12}$   \\
$kT_{blue}$\,(keV)                     &  $(1.08\pm0.08)\times10^3$      \\
$z$                                    &  $2.67 \pm 0.37$                \\
\hline
$\langle n_{CBM} \rangle$ (cm$^{-3}$)  &  $(4.7\pm1.2)\times10^{-5}$   \\
\hline
\end{tabular}
\caption{The results of the simulation of GRB 090227B in the fireshell model.}
\label{tab:2}
\end{table}

We now proceed with the analysis of the subsequent emission to derive the properties of the surrounding CBM. 

\subsection{Analysis of the prompt emission}\label{sec:4.2}

\begin{figure}
\centering
\includegraphics[width=\hsize,clip]{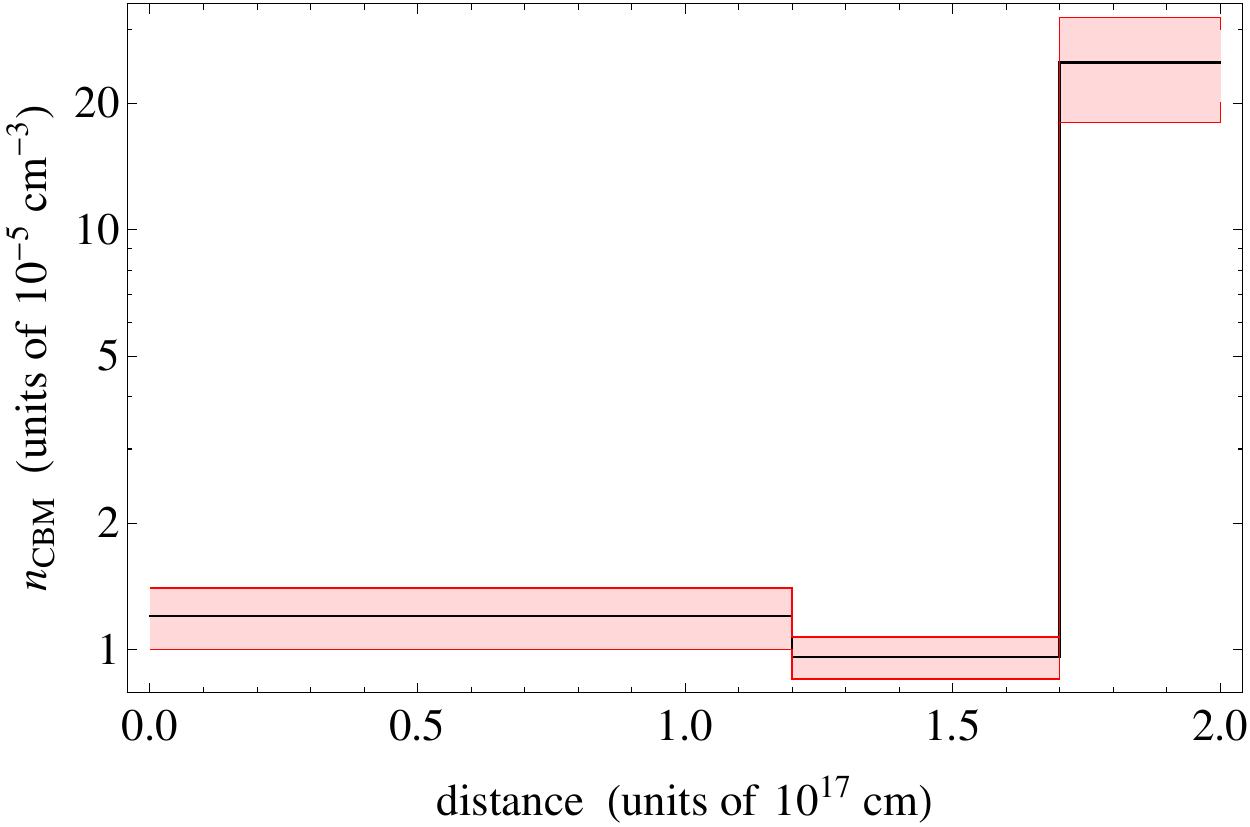} 
\caption{The radial CBM number density distribution of GRB 140619B (black line) and its range of validity (red shaded region).}
\label{fig:5}
\end{figure}

\begin{figure}
\centering
\includegraphics[width=\hsize,clip]{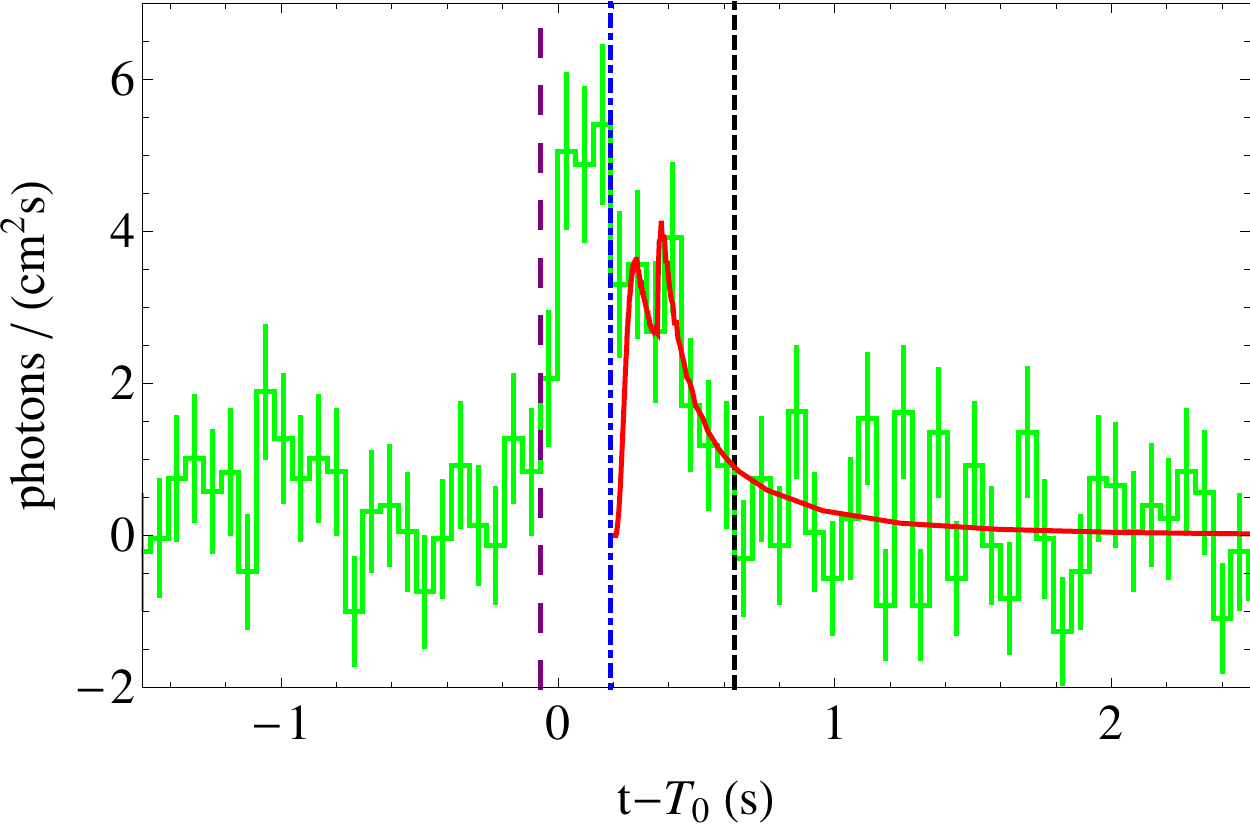}
\caption{The BGO-b1 ($260$ keV -- $40$ MeV) simulated light curve of the prompt emission of GRB 140619B (solid red line). Each spike corresponds to the CBM number density profile described in Tab.~\ref{tab:2} and Fig.~\ref{fig:5}. The blue dot-dashed vertical line marks the end of the P-GRB emission. The purple long-dashed and the black dashed vertical lines indicate, respectively, the starting and the ending times of the $T_{90}$ time interval. 
Clearly visible outside of this time interval is the background noise level. The continuation of the simulation after $T_{90}$ is due to the residual large angle emission of the EQTS \citep{Bianco2005b,Bianco2005a} due to the density profile indicated in Tab.~\ref{tab:3}.}
\label{fig:6}
\end{figure}

Having determined the initial conditions for the fireshell, i.e., $E^{tot}_{e^+e^-}=6.03\times10^{52}$ ergs and $B=5.52\times10^{-5}$, the dynamics of the system is uniquely established.
In particular, we obtain the Lorentz factor at transparency, $\Gamma_{tr}=1.08\times10^4$, and we can simulate the light curve and the spectrum of the prompt emission.
To reproduce the pulses observed especially in the BGO-b1 light curve (see Fig.~\ref{fig:1}) we have derived the radial distributions of the CBM number density and of the filling factor $\mathcal{R}$ around the burst site (see Tab.~\ref{tab:3} and Fig.~\ref{fig:5}).
The errors in the CBM number density and in $\mathcal{R}$ are defined as the maximum possible variation of the parameters to guarantee agreement between the simulated light curve and the observed data.
The final simulation of the BGO-b1 light curve ($260$ keV -- $40$ MeV) is shown in Fig.~\ref{fig:6}. 
\begin{table*}
\centering
\begin{tabular}{cccccc}
\hline\hline
Cloud     &  Distance (cm)        &  $n_{CBM}$ (cm$^{-3}$)       &  $\mathcal{R}$   &  $\Gamma$  &  $d_v$ (cm)               \\
\hline
$1^{th}$  &  $1.50\times10^{15}$  &  $(1.2\pm0.2)\times10^{-5}$  &  $(2.8\pm0.3)\times10^{-11}$  &  $1.08\times10^4$  &  $2.76\times10^{11}$  \\
$2^{nd}$  &  $1.20\times10^{17}$  &  $(9.2\pm1.1)\times10^{-6}$  &  &  $2.07\times10^3$   &  $1.16\times10^{14}$  \\
\hline
$3^{rd}$  &  $1.70\times10^{17}$  &  $(2.5\pm0.5)\times10^{-4}$  &  $(3.5\pm0.6)\times10^{-10}$ &  $1.84\times10^3$   &  $1.85\times10^{14}$ \\
\hline
\end{tabular}
\caption{The density and filling factor masks of GRB 140619B. In each column are listed, respectively, the CBM cloud, the corresponding initial radius away from the BH, the number density, the filling factor, the Lorentz factor, and the total transversal size of the fireshell visible area.}
\label{tab:3}
\end{table*}

Interestingly, the average CBM number density in GRB 140619B, $\langle n_{CBM} \rangle = (4.7\pm1.2)\times10^{-5}$ cm$^{-3}$ (see Tab.~\ref{tab:3}), is very similar to that of the prototype GRB 090227B, $\langle n_{CBM} \rangle = (1.90\pm0.20)\times10^{-5}$ cm$^{-3}$.
In both the cases the CBM densities are typical of the galactic halo environment.

\begin{figure}
\centering
\includegraphics[width=\hsize,clip]{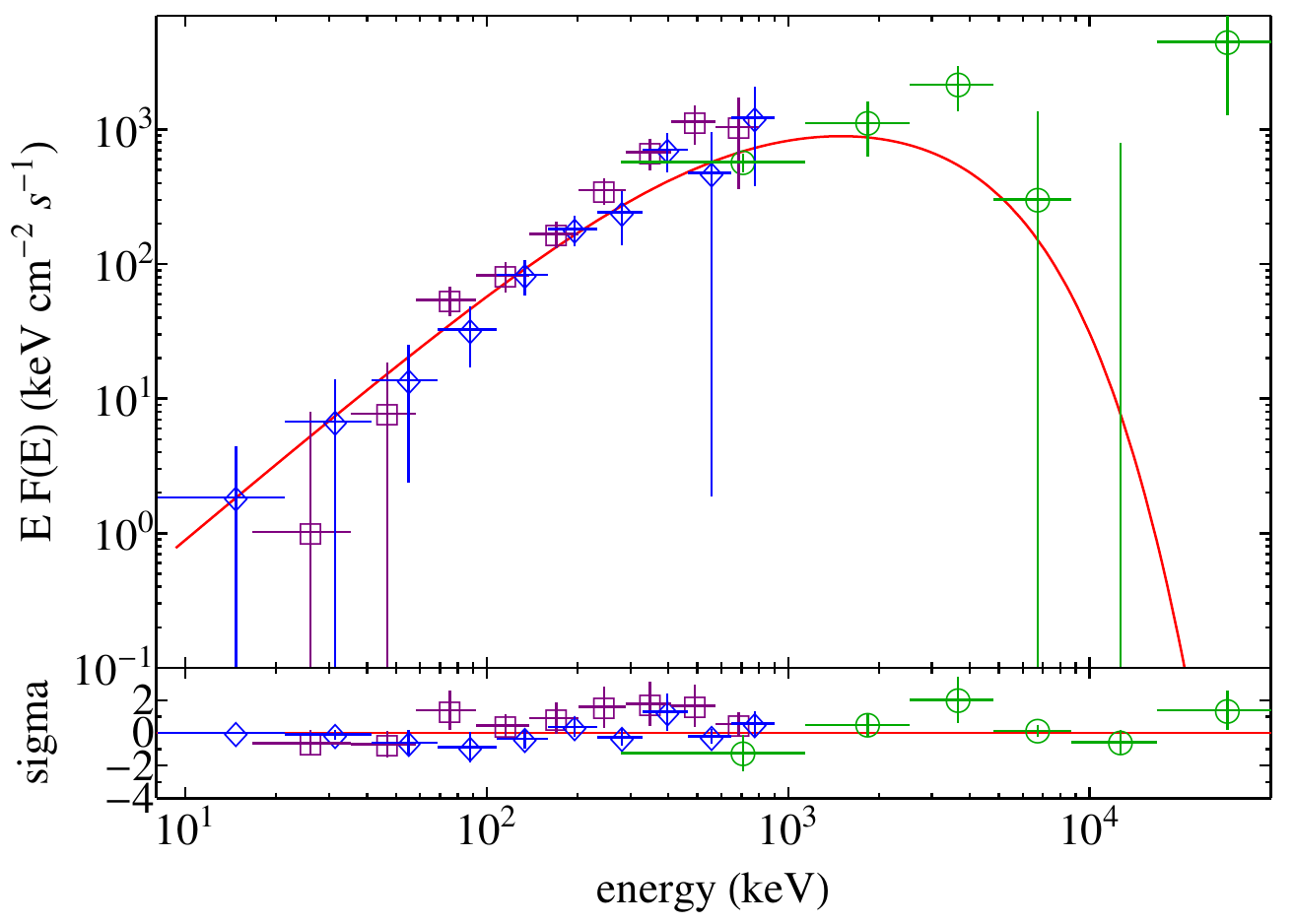}
\caption{Top panel: comparison between the $8$--$900$ keV data from the NaI-n6 (purple squares) and n9 (blue diamonds) detectors, and the $260$ keV -- $40$ MeV data from the BGO-b1 detector (green circles), and the simulation within the firshell model (solid red curve) in the time interval $\Delta T_2$. Bottom panel: the residuals of the above mentioned data with the simulation.}
\label{fig:7}
\end{figure}

We turn now to the spectrum of the prompt emission using the spectral model described in \citet{Patricelli} with a phenomenological parameter $\alpha=-1.11$. 
From fitting the light curve in the energy range $260$ keV -- $40$ MeV, we have extended the simulation of the corresponding spectrum down to $8$ keV to check overall agreement with the observed data.
The final result is plotted in Fig.~\ref{fig:7}, where the rebinned NaI-n6 and n9 and BGO-b1 data in the $\Delta T_2$ time interval show their agreement with the simulation; the lower panel in Fig.~\ref{fig:7} shows the residuals of the data around the fireshell simulated spectrum.

The fireshell approach is different from the fireball one, where the sharp luminosity variations observed in the prompt emission are attributed to the prolonged and variable activity of the ``inner engine'' \citep[see, e.g.][]{ReesMeszaros1994,Ramirez2000,Piran2004P}.

In the fireshell model, the observed time variability of the prompt emission is produced by the interaction of the accelerated baryons of the fireshell with the CBM ``clumps '' \citep[see, e.g.,][]{Ruffini2002,Ruffini2006,Patricelli}. The issue of the time variability in GRB light curves has been long debated. \citet{Zhang2006} and \citet{NakarGranot2007} indicated difficulties in producing short time variability from CBM inhomogeneities. The opposite point of view has been expressed by \citet{DermerMitman1999} and \citet{Dermer2006,Dermer2008}. In the fireshell model it has been shown that, from the correct computation of the equations of motion of the shell, of the EQTS, and of the Lorentz factor \citep[][and Sec.~\ref{sec:2}]{Bianco2005b,Bianco2005a}, the short time scale variability of GRB light curves occurs in regimes with the larger values of the Lorentz factor, when the total visible area of the emission region is very small and ``dispersion'' in arrival time of the luminosity peaks is negligible. 
Therefore the short time scale variability indeed can be produced by the CBM inhomogeneities \citep[see Sec.~3 in][]{Patricelli}. 
This has been verified in the present case of GRB 140619B, where the values of the Lorentz factor $\Gamma$ and the total transversal size of the fireshell visible area $d_v$ at the initial radius of the CBM cloud are explicitly indicated in Tab.~\ref{tab:3}. 
These values of $d_v$ are smaller than the thickness of the inhomogeneities ($\Delta r\approx10^{16}$--$10^{17}$ cm) and fully justify the adopted spherical symmetry approximation \citep{Ruffini2002,Ruffini2006,Patricelli}. Consequently, a finer description of each substructure in the spikes observed in the light curve is not necessary and does not change the substantial agreement of the model with the observational data, which is provided by the average densities and the filling factors in Tab.~\ref{tab:3}.

\subsection{The progenitor system}\label{sec:4.3}

In analogy with the case of GRB 090227B \citep[see, e.g.,][]{Oliveira2014, Muccino2013}, we conclude that the progenitor of GRB 140619B is a NS--NS merger.
As a lower limit, we have considered the simplest case by assuming two NSs with the same mass $M_{NS}$ such that the total mass would be larger than the neutron star critical mass $M_{crit}^{NS}$, e.g. $2M_{NS}\gtrsim M_{crit}^{NS}$.
This condition is clearly necessary for the formation of a BH and the consequent application of the fireshell model.
It is also appropriate here to recall that only a subset of binary NSs mergers can fulfill this stringent requirement (see Fig.~\ref{fig:dns}).
This will strongly affect the estimate of the rate of these \familytwo short GRBs, when compared with the usual expected binary NS rate (see Sec.~\ref{sec:4.5} and Conclusions).

Referring to the work of \citet{Belvedere} on nonrotating NSs in the global charge neutrality treatment with all the fundamental interactions taken into account properly, we have considered two NSs with mass $M_{NS}=1.34$ M$_\odot = 0.5 M_{crit}^{NS}$ and corresponding radius $R=12.24$ km.
As a working hypothesis we assume that in the NS merger the crustal material from both NSs contributes to the GRB baryon load, while the NS cores collapse to a BH.
For each NS the crustal mass from the NL3 nuclear model is $M_c=3.63\times10^{-5}$ M$_\odot$,
so the total NS merger crustal mass is $M_{2c}=2M_c=7.26\times10^{-5}$ M$_\odot$.
On the other hand, the baryonic mass engulfed by the $e^+e^-$ plasma before transparency is $M_B=E^{tot}_{e^+e^-}B/c^2=(1.86\pm0.35)\times10^{-6}$ M$_\odot$,
so we can conclude that only a small fraction of the crustal mass contributes to the baryon load, namely $M_B=(2.56\pm0.48)\%M_{2c}$.
This value is consistent with the global charge neutrality condition adopted in \citet{Belvedere}.
The usually adopted LCN condition leads instead to a crustal mass $M^{LCN}_c\sim 0.2~M_\odot$ \citep[see, e.g.,][]{Belvedere,Oliveira2014}, which would be inconsistent with the small value of the baryon load inferred above.

\section{On the GWs emission and the detectability or absence thereof}\label{GWs}

Following the previous work on GRB 090227B \citep{Oliveira2014}, we now estimate the emission of GWs of the binary NS progenitor of the short GRB 140619B using the effective-one-body (EOB) formalism \citep{1999PhRvD..59h4006B,2000PhRvD..62f4015B,2000PhRvD..62h4011D,2001PhRvD..64l4013D,2010PhRvD..81h4016D} and assess the detectability of the emission by the Advanced LIGO interferometer%
.\footnote{http://www.advancedligo.mit.edu} 
The EOB formalism maps the conservative dynamics of a binary system of nonspinning objects onto the geodesic dynamics of a single body of reduced mass $\mu=M_1 M_2/M$, with total binary mass $M=M_1+M_2$. The effective metric is a modified Schwarzschild metric with a rescaled radial coordinate, $r = c^2 r_{12}/(G M)$, where $r_{12}$ is the distance between the two stars. The binary binding energy as a function of the orbital frequency $\Omega$  is given by $E_b(\Omega) = M c^2 [\sqrt{1 + 2 \nu (\hat{H}_{\rm eff} - 1})-1 ]$, where the effective Hamiltonian 
$ \hat{H}_{\rm eff}^2  = A(u) + p_{\phi}^2 B(u)$
depends on the radial potential $A(u)$ of the variable $u=1/r$ and  $B(u) = u^2 A(u)$, while the angular momentum for the circular orbit is given by $ p_{\phi}^2 = - A'(u)/[u^2 A(u)]'$, where a prime stands for the derivative with respect to $u$ \citep[see, e.g.,][for further details]{2013PhRvD..87l1501B}. 
In order to obtain the derivative of the effective Hamiltonian $\hat{H}_{\rm eff}$ as a function of $\Omega$,  we must use the chain rule together with the relation $\Omega=\Omega(u)$ following from the angular Hamilton equation of motion in the circular case $G M\Omega(u) = (1/u) \partial H/\partial p_{\phi} = M A(u) p_{\phi}(u) u^2/(H \hat{H}_{\rm eff})$, where $G$ is the gravitational constant.
Finally we obtain the rate of orbital energy loss through emission of GWs from the related derivative $dE_b/d\Omega$.

\begin{figure}\label{amplitude}
\centering
\includegraphics[width=\hsize,clip]{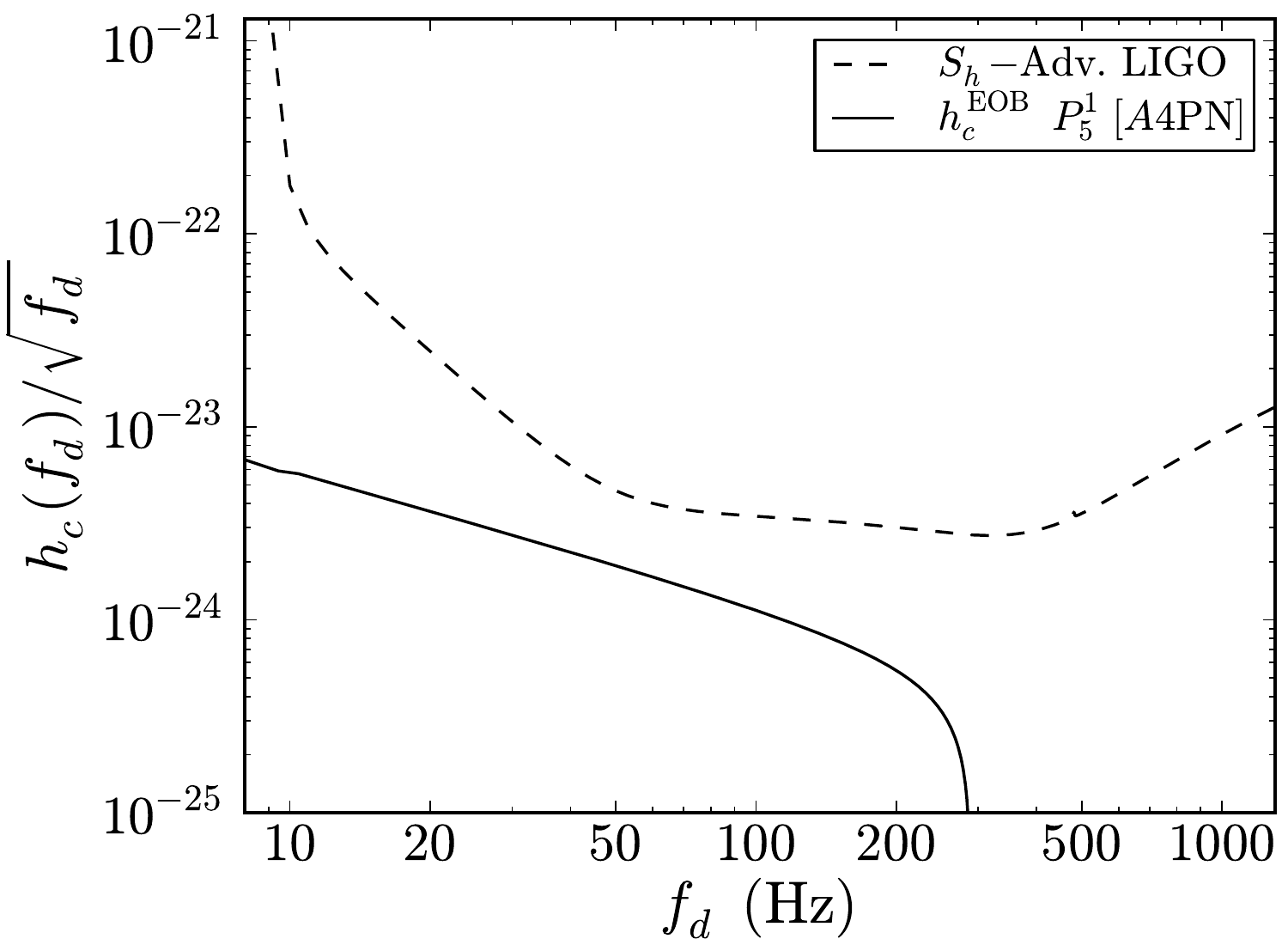}
\caption{The sensitivity curve of Advanced LIGO $S_h(f)$ (dashed black curve) and the characteristic gravitational amplitude $h_c(f_d)/\sqrt{f_d}$ (solid black curve) of the binary NS progenitor of GRB 140619B, as a function of the frequency at the detector $f_d$. The EOB radial potential $A(u)$ was calculated using values for the coefficients in the 4th order post-Newtonian (PN) approximation and $P^1_5$ is the Pad\`e approximant of order (1,5).}
\end{figure}

Using the well known matched filtering technique, we compute the signal-to-noise ratio (SNR) from the Fourier transform of the signal $h(t)=h_+F_{+} + h_{\times} F_{\times}$, where $h_{+,\times}$ are functions that depend on the direction and polarization of the source and $F_{+,\times}$ depend on the direction of the detector. By making an rms average over all possible source directions and wave polarizations, i.e., $\langle F^2_+\rangle =\langle F^2_{\times}\rangle = 1/5$, we obtain \citep[see][for details]{1998PhRvD..57.4535F}
\begin{equation}\label{eq:SNR}
\langle{\rm SNR}^2\rangle=\int_{f_{\rm min}}^{f_{\rm max}} d f_d \frac{h^2_c(f_d)}{5 f^2_d S^2_h(f_d)},
\end{equation}
where $S_h(f)$ is the strain noise spectral density (in units 1/$\sqrt{{\rm Hz}}$) of the interferometer. We have also introduced the characteristic GW amplitude, $h_c$, defined using the Fourier transform of the GW form $h(t)$, $h_c(f)=f|\tilde{h}(f)|$, and it is given by
\begin{equation}\label{hc}
 h_c^2(f)= \frac{2(1+z)^2}{\pi^2 d_L^2} \frac{d E_b}{d f} [(1+z) f_d],
\end{equation}
with $z$ the cosmological redshift, $f_d= f/(1+z)$ the GW frequency at the detector, $f=\Omega/\pi$ the frequency in the source frame, $f_{\rm min}$ the minimal bandwidth frequency of the detector, and $f_{\rm max}=f_c/(1+z)$ the maximal bandwidth frequency, where $f_c=\Omega_c/\pi$ is the binary contact frequency and $d_L$ is the luminosity distance. In Fig.~\ref{amplitude} we show the strain-noise sensitivity of Advanced LIGO, $S_h(f)$, and the characteristic gravitational amplitude per square root frequency, $h_c(f_d)/\sqrt{f_d}$, both plotted as functions of the frequency at the detector $f_d$. 

Following the above procedure we obtained for the short GRB 140619B a very low value $\langle{\rm SNR}\rangle \approx 0.21$ compared to the value $\rm SNR=8$ needed for an optimal positive detection. The low value of the SNR is clearly due to the large cosmological distance to the source, $d\approx21$~Gpc. Although the rms-averaged SNR we have computed might improve by a factor $\approx 5/2$ for an optimally located and polarized source (e.g.~$\langle F^2_+\rangle = 1$ and $\langle F^2_{\times}\rangle=0$) with an optimal face-on orbit ($\cos\iota = 1$), in the case of GRB 140619B it would increase only to a maximal value ${\rm SNR(opt)} \approx 0.5$. From the dynamics of the system, we also find that this binary emits a total energy of $E^T_{\rm GW}=7.42\times10^{52}$~erg  in gravitational radiation during the entire inspiral phase all the way up to the merger.

\section{Considerations on the GeV emission of GRB 140619B}\label{sec:4.4}

\begin{figure}
\centering
\includegraphics[width=\hsize,clip]{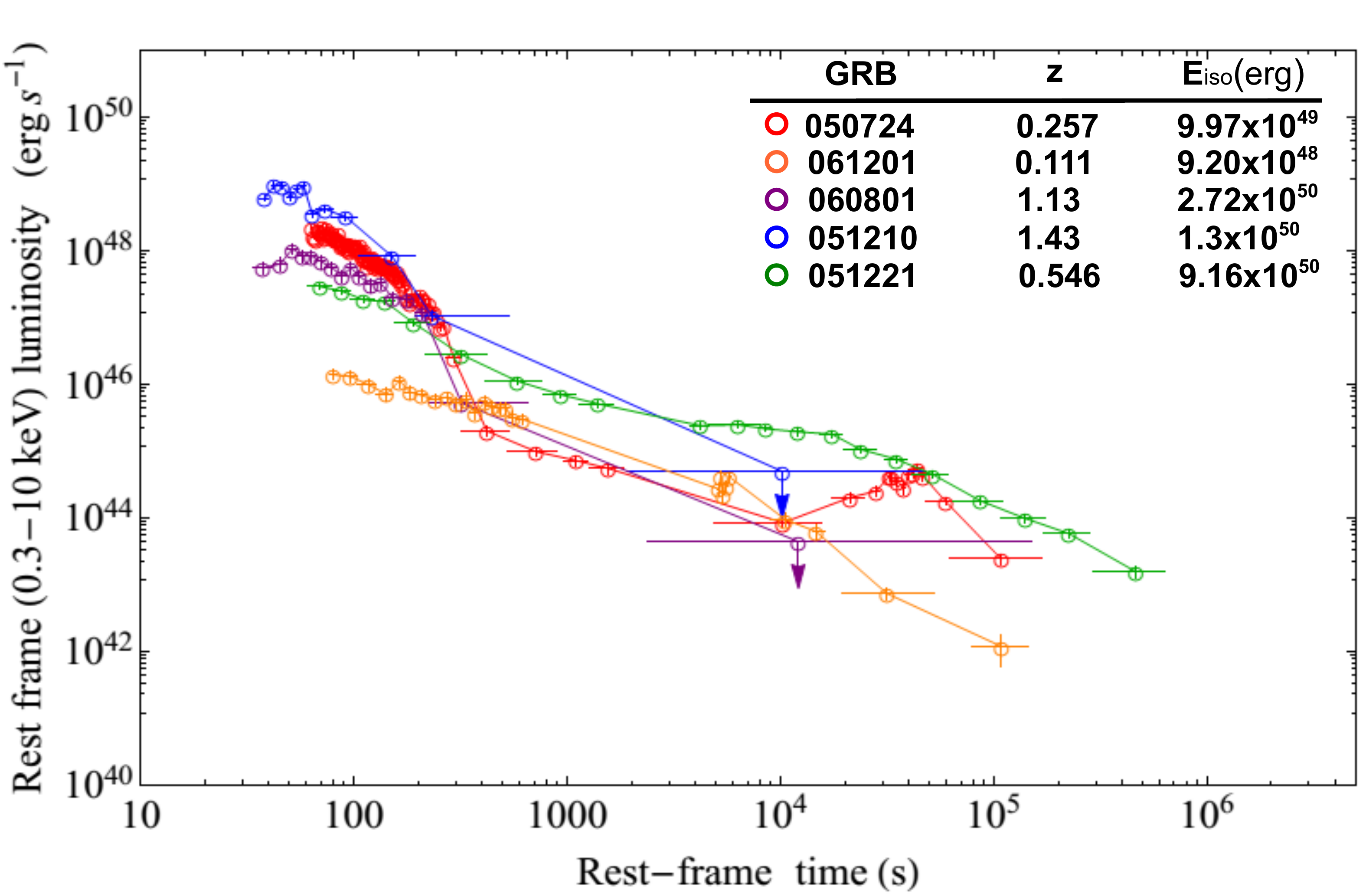}
\includegraphics[width=\hsize,clip]{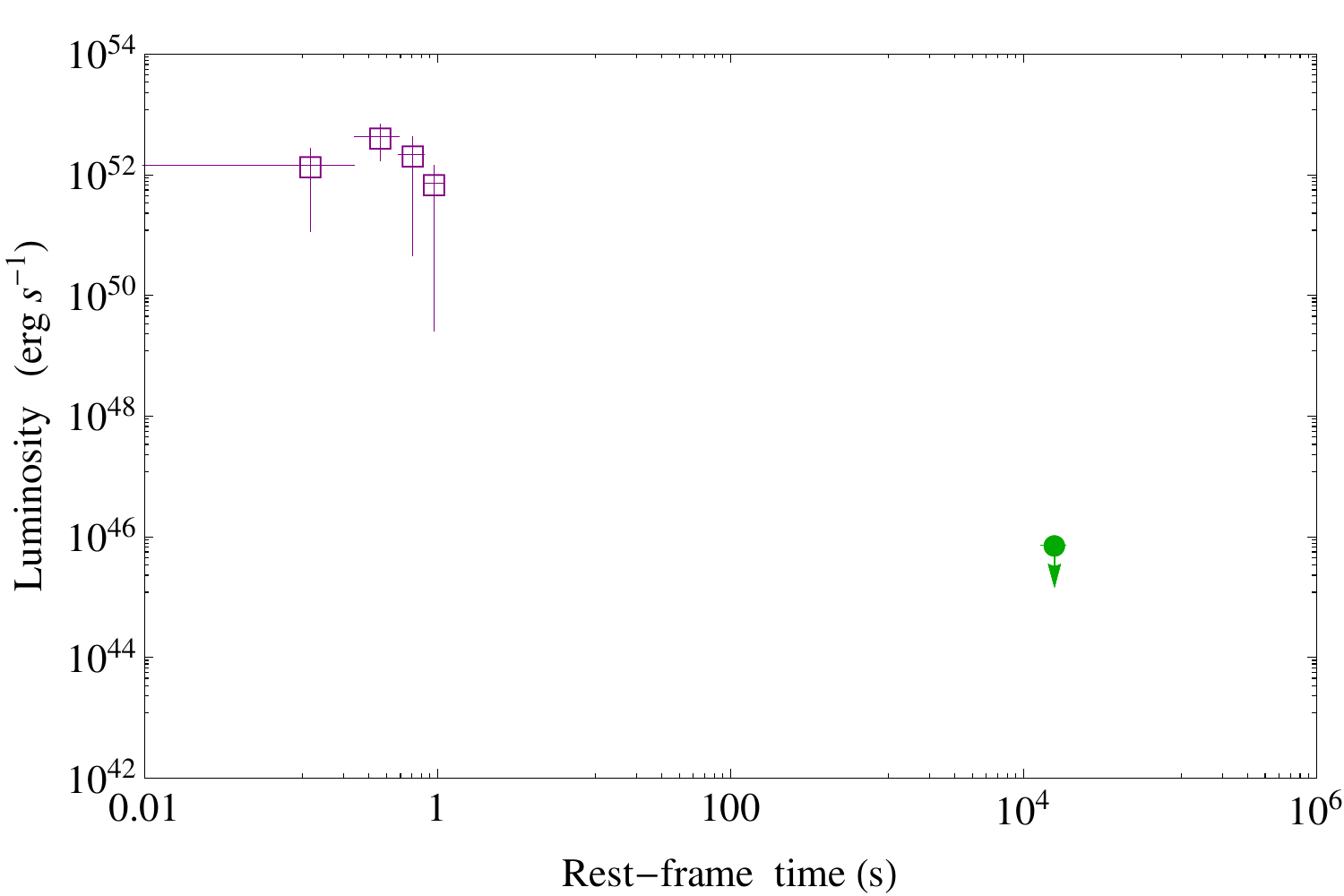}
\includegraphics[width=\hsize,clip]{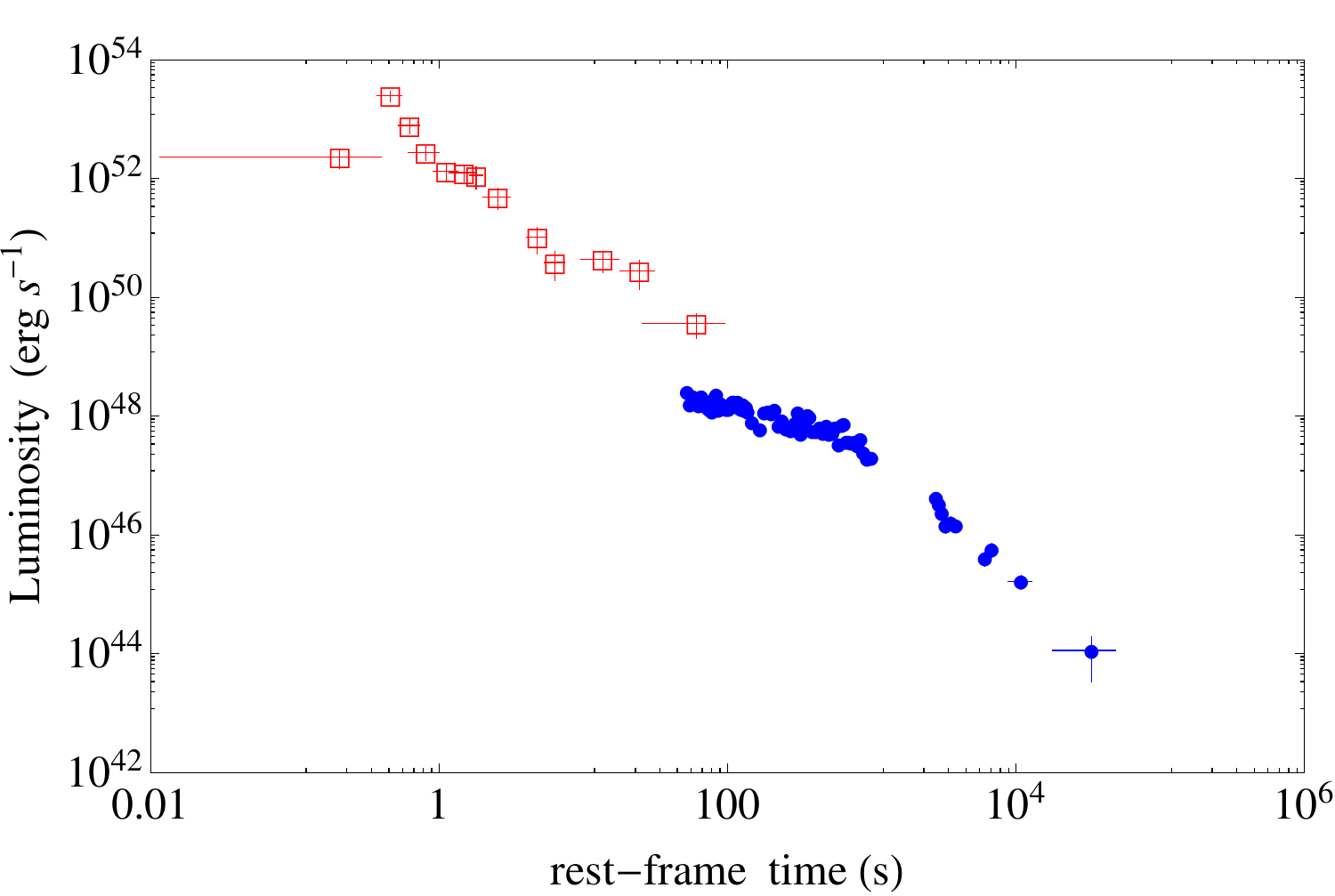}
\caption{Top panel (a): the rebinned rest-frame $0.3$--$10$ keV X-ray luminosities of weak short GRBs leading to massive NSs; the corresponding bursts, redshifts and energies are indicated in the legend. In their afterglows there is no regular power-law behavior at late times and no nesting \citep{Ruffini2014}.
Middle panel (b): the short lived rest-frame $0.1$--$100$ GeV isotropic luminosity light curve (purple squares) and the rest-frame $0.3$--$10$ keV upper limit, as set from the analysis of GRB 140619B outlined in Sec.~\ref{sec:2} (green circle). Bottom panel (c): the long lived rest-frame $0.1$--$100$ GeV (red squares) and the rest-frame $0.3$--$10$ keV (blue circles) isotropic luminosity light curves of GRB 090510.}
\label{fig:8}
\end{figure}

In addition to the analogies with GRB 090227B, GRB 140619B presents a novelty of special interest: a short-lived emission ($\sim4$ s) observed at energies $\gtrsim0.1$ GeV.
The light curve of this emission shows a rising part which peaks at $\sim2$ s, followed by a decaying tail emission lasting another $\sim2$ s in the observer frame (see Fig.~\ref{fig:8}b).
Since GRB 140619B was in the LAT FoV during the entire observational period, the absence of emission after $\sim4$ s has been attributed to a cut-off intrinsic to the source. 
We divided the overall emission into four time intervals (see Fig.~\ref{fig:8}b), each of them lasting $1$ s. The corresponding spectra are best fit by power-law models.
The total isotropic energy of the $0.1$--$100$ GeV emission is $E_{LAT}=(2.34\pm0.91)\times10^{52}$ erg.

In complete analogy with the GeV emission emitted in the binary-driven hypernovae (BdHNe), we attribute this high energy radiation to the newly-formed BH.
This identification is clearer here in view of the absence of a supernova (SN) and the related constant power-law emission in X-rays, when measured in the cosmological rest-frame of the BdHN \citep{Ruffini2014,Yu2014,RR2CL}.

The presence of this GeV emission is not a peculiarity of GRB 140619B, but is a common feature of all these \familytwo short GRBs.
In line with this, the apparent absence of the GeV emission in GRB 090227B has already been discussed in Sec.~\ref{sec:3}: it can be explained simply by the fact that this source was outside the nominal LAT FoV. 
The significant detection in the LAT low energy channel and the presence of only one transient-class event with energy above $100$ MeV associated with the GRB \citep{Ackermann2013} confirms that in optimal conditions the GeV emission from GRB 090227B should have been detected.

Now consider GRB 090510, which has the characteristics of the \familytwo short GRBs ($E_{iso}>10^{52}$ erg and $E_{p,i}>2$ MeV), including the presence of a high energy GeV emission lasting $\sim10^2$ s.
This high energy emission continues up to the signal goes below the LAT threshold \citep{Ackermann2013}. 
The new feature of GRB 090510, among the \familytwo short GRBs, is a well determined cosmological redshift inferred from the optical observations. 
The corresponding distance indeed coincides with the one theoretically predicted in the fireshell binary merger model (Muccino et al., in preparation).

In Fig.~\ref{fig:8}a we compare and contrast the afterglows of the traditional low energetic short GRBs \citep[see][for a review]{2014ARA&A..52...43B} with those of the \familytwo short GRB 140619B (see Fig.~\ref{fig:8}b) and GRB 090510 (see Fig.~\ref{fig:8}c).
In Fig.~\ref{fig:diagram} we show the evolution of the NS--NS merger generating a \familytwo short GRB.
In this system the conservation laws for total energy and the total angular momentum have to be satisfied during and following the binary NS merger (Rueda et al., in preparation). 
One of the most important issues is the determination of the dimensionless angular momentum $c\,J/(GM^2)$ of the newly-born BH (where $J$ and $M$ are, respectively, the BH spin angular momentum and mass). 
These considerations have been applied to GRB 090510 (Muccino et al., in preparation).

Before closing, we call attention to GRB 081024B, which we are currently addressing within the fireshell model (Aimuratov et al., in preparation), and which shows all the typical features of the \familytwo short GRBs, including a distinctive GeV emission. 
In conclusion, we can safely assert that all \familytwo short GRBs, when the observational requirements are fulfilled, present a short-lived but very intense GeV emission, which in our interpretation originates from the newly-formed BH.

In Tab.~\ref{tab:LAT} we listed the redshift, $E_{p,i}$, $E_{iso}$, and the GeV isotropic emission energy $E_{LAT}$ in the rest-frame energy band $0.1$--$100$ GeV of the three \familytwo short GRBs discussed here.
In computating $E_{iso}$ we have inserted the energy computed in the rest-frame energy band $1$--$10000$ keV.
\begin{table}
\centering
\begin{tabular}{ccccc}
\hline\hline
GRB         &  z                            &  $E_{p,i}$          &    $E_{iso}$              &   $E_{LAT}$           \\
               &                               &  (MeV)                &    ($10^{52}$ erg)    &    ($10^{52}$ erg) \\
\hline
081024B  &  $>3.0$                  &  $>8.2$              &    $>2.4$                  &  $>2.7$      \\
090227B  &  $1.61\pm0.14$       &  $5.89\pm0.30$  &  $28.3\pm1.5$          &    -                        \\
090510    &  $0.903\pm0.003$   &  $7.89\pm0.76$  &  $3.95\pm0.21$         &  $5.78\pm0.60$    \\
140619B  &  $2.67\pm0.37$       &  $5.34\pm0.79$  &  $6.03\pm0.79$        &   $2.34\pm0.91$      \\
\hline
\end{tabular}
\caption{The redshift, the rest-frame peak spectral energy, the isotropic energy $E_{iso}$ in the rest-frame energy band $1$--$10000$ keV, and the GeV isotropic  emission energy $E_{LAT}$ in the rest-frame energy band $0.1$--$100$ GeV of the four \familytwo short GRBs discussed here. The values indicated for GRB 081024B will be discussed in Aimuratov et al. (in preparation).}
\label{tab:LAT}
\end{table}

\section{The rate of family-2 short GRBs}\label{sec:4.5}

With the identification of three \familytwo short GRBs, namely GRB 090227B and GRB 140619B, with theoretically inferred redshifts, and GRB 090510 with a measured redshift, all of them detected by the \textit{Fermi} satellite, we are now in a position to give an estimate of the expected rate $\rho_0$ of such events. 
Following \citet{Soderberg2006Nature} and \citet{Guetta2007}, for these sources we have computed the $1$ s peak photon flux $f_p$ in the energy band $1$--$1000$ keV, which is $16.98$ photons cm$^{-2}$s$^{-1}$ for GRB 090227B, $9.10$ photons cm$^{-2}$s$^{-1}$ for GRB 090510, and $4.97$ photons cm$^{-2}$s$^{-1}$ for GRB 140619B.
From the spectral parameters for each source, we have computed $f_p$
for various redshifts 
until it coincided with the corresponding threshold peak flux $f_T$ which is the limiting peak photon flux  allowing burst detection \citep[see the analysis in][for details]{Band2003}.
In this way we have evaluated for each source the maximum redshift $z_{max}$ at which the burst would have been detected and, then, the corresponding maximum comoving volume $V_{max}$.
For GRB 140619B we obtain $f_p\equiv f_T=1.03$ photons cm$^{-2}$s$^{-1}$ at maximum redshift $z_{140619B}^{max}=5.49$; for GRB 090227B, which is the brightest one, we find $f_p\equiv f_T=1.68$ photons cm$^{-2}$s$^{-1}$ at a maximum redshift $z_{090227B}^{max}=5.78$; finally, for GRB 090510, we get $f_p\equiv f_T=1.96$ photons cm$^{-2}$s$^{-1}$ at a maximum redshift $z_{090510}^{max}=2.25$. 
Correspondingly we have computed $V_{max}$.

The empirical rate can be evaluated as 
\begin{equation}
\label{rate}
\rho_0=\left(\frac{\Omega_F}{4\pi}\right)^{-1}\frac{N}{V_{max} T_F}\ ,
\end{equation}
where $N=3$ is the number of identified energetic NS--NS short bursts, $\Omega_F \approx 9.6$ sr is the average Fermi solid angle, and $T=6$ years is the Fermi observational period.  
We infer a local rate of $\rho_0=\left(2.6^{+4.1}_{-1.9}\right)\times10^{-4}$Gpc$^{-3}$yr$^{-1}$, where the attached errors are determined from the $95\%$ confidence level of the Poisson statistic \citep{Gehrels1986}.
At $z\geq0.9$, the above inferred rate provides an expected number of events $N_>=4^{+6}_{-3}$, which is consistent with the above three observed events during the Fermi observational period. Also at $z\leq0.9$ our estimate $N_<=0.2^{+0.31}_{-0.14}$ is consistent with the absence of any \familytwo short GRB detection.

With the inclusion of GRB 081024B, with a theoretically estimated redshift $z>3$ (more details will appear in Aimuratov et al., in preparation), the above rate remains stable with smaller error bars, i.e., $\rho_0=\left(2.1^{+2.8}_{-1.4}\right)\times10^{-4}$Gpc$^{-3}$yr$^{-1}$.
This inferred rate is different from that of the long GRBs, recently estimated to be $\rho_{L-GRB}=1.3^{+0.7}_{-0.6}$ Gpc$^{-3}$yr$^{-1}$ \citep{Wanderman2010}, and also from the estimates of the family-1 short GRBs given in the literature (without a beaming correction $\rho_{short}=1$--$10$ Gpc$^{-3}$yr$^{-1}$, see e.g., \citealp{2014ARA&A..52...43B} and \citealp{2014arXiv1409.8149C}).

Such a low rate can be explained based upon the existing data of binary NSs within our Galaxy (see Sec.~\ref{sec:NS}). From Fig.~\ref{fig:dns} we notice that only a subset of them has the sum of the masses of the components larger than the critical NS mass and can collapse to a BH in their merger process.
Only this subset can lead to a \familytwo short GRB.

\section{The family-2 short GRBs and the $E_{p,i}$--$E_{iso}$ relation for short GRBs}\label{sec:Calderone}

Now we discuss some general considerations for the new $E_{p,i}$--$E_{iso}$ relation for short GRBs \citep{2012ApJ...750...88Z,Calderone2014}, with a power law similar to the one of the Amati relation for long GRBs \citep{Amati2008}, but different amplitude. 
This yet unexplained difference discourages the use of the Amati relation as an astronometrical tool.
All four family-2 short GRBs satisfy this new $E_{p,i}$--$E_{iso}$ relation (see the quantities listed in Tab.~\ref{tab:LAT}).
We call attention to the need to investigate the physical reasons for the validity of this universal $E_{p,i}$--$E_{iso}$ relation, which appears to be satisfied by \familyone short bursts, where the binary NS merger does not lead to BH formation, and also the \familytwo short bursts, where BHs are formed and reveal their presence by giving rise to the short-lived but significant GeV emission.

\section{Conclusions}

In this article we have predicted the occurrence of two different kinds of short GRBs originating from binary NS mergers, based on 
\begin{itemize}
\item[a)] the analysis of GRB 090227B, the prototype of short bursts originating from a binary NS leading to BH formation \citep{Muccino2013}, 
\item[b)] the recent progress in the determination of the mass-radius relation of NSs and the determination of their critical mass $M_{crit}^{NS}\approx2.67$ M$_{\odot}$ \citep{Rotondo2011,Rueda2011,Belvedere,Belvedere2014,Rueda2014,BelRR}, and
\item[c)] the recently measured mass of PSR J0348+0432, $M=(2.01\pm0.04)$ M$_\odot$ \citep{2013Sci...340..448A}, establishing an absolute lower limit on $M_{crit}^{NS}$, and the remarkable information gained from radio observations of binary NS systems in our own Galaxy \citep{2011A&A...527A..83Z,2014arXiv1407.3404A}. 
\end{itemize}

The first kind of short GRBs, which we call \familyone, are the most common ones with $E_{iso}<10^{52}$ erg and rest-frame spectral peak energy $E_{p,i}<2$ MeV, originating from binary NS mergers with merged core mass smaller than $M_{crit}^{NS}$ and leading, therefore, to a massive NS, possibly with a companion. We identify these \familyone short bursts with the ones extensively quoted in literature \citep[see, e.g.,][for a review]{2014ARA&A..52...43B}.

The second kind of short GRBs, which we call \familytwo, are those with $E_{iso}>10^{52}$ erg and harder spectra with $E_{p,i}>2$ MeV, 
originating from binary NS mergers with merged core mass larger than $M_{crit}^{NS}$. These \familytwo short bursts satisfy the necessary condition to form a BH, following the example of the prototype GRB 090227B \citep{Muccino2013}.

The application of the fireshell model \citep{Ruffini2001c,Ruffini2001,Ruffini2001a} to the \familytwo short GRB 140619B analyzed here has allowed the determination of the physical parameters of this source: the identification of the P-GRB emission in the early $\sim0.2$ s of its light curve, the theoretical cosmological redshift of $z=2.67\pm0.37$ and consequently the total burst energy $E^{tot}_{e^+e^-} = (6.03\pm0.79)\times10^{52}$ ergs, the baryon load $B = (5.52\pm0.73)\times10^{-5}$, and a Lorentz $\Gamma$ factor at transparency $\Gamma_{tr} = (1.08\pm0.08)\times10^4$.  The analysis of the prompt emission has also led to the determination of the CBM density, $\langle n_{CBM}\rangle = (4.7\pm1.2)\times10^{-5}$ cm$^{-3}$, typical of the galactic halo environment, where NS--NS binaries migrate to, due to natal kicks imparted to them at birth \citep[see, e.g.,][]{2014ARA&A..52...43B,Narayan1992,1999MNRAS.305..763B,1999ApJ...526..152F,2006ApJ...648.1110B}, clearly supporting the binary NS merger hypothesis of this source. Unexpectedly, we have found the existence of a short-lived and very intense GeV emission, just after the P-GRB occurrence and during and after the prompt emission phase, which has led us to conclude that this high energy emission originates from the newly-formed BH.

While this article was being refereed, we have discovered three additional examples of these \familytwo short bursts: GRB 081024B, GRB 090510, and GRB 090227B. These have given evidence  that all these \familytwo short bursts indeed show the existence of high energy emission, with the sole exception of GRB 090227B, which at the time of the observation was outside the nominal LAT field of view.

In summary we formulate some norms and theoretical predictions.
\begin{itemize}
\item[1)] All \familyone short GRBs have an extended X-ray afterglow \citep[see, e.g., Fig.~\ref{fig:8}a and][]{2014ARA&A..52...43B}. When computed in the rest-frame $0.3$--$10$ keV energy band they do not show any specific power-law behavior \citep{Pisani2013} or the ``nesting'' properties \citep{Ruffini2014} which have been discovered in some long GRBs. We predict that \familyone short GRBs, originating from a binary merger to a massive NS, should never exhibit high energy emission. The upper limit of $10^{52}$ erg can be simply understood in terms of a merger leading to a massive NS.
\item[2)] All \familytwo short GRBs have been observed not to have prominent X-ray or optical afterglows. They all have short-lived but very energetic GeV emissions (see, e.g., Fig.~\ref{fig:8}b and c), when LAT data are available. 
The upper limit of $10^{54}$ erg can be also simply understood in terms of a merger leading to BH formation.
\item[3)] The high energy emission episode in \familytwo short GRBs always occurs at the end of the P-GRB emission, during and after the prompt emission phase. This fact uniquely links the high energy emission to the occurrence of the newly-born BH. The prompt emission phase studied within the fireshell model has also allowed the determination of a large number of essential astrophysical parameters, both of the source (e.g., $E^{tot}_{e^+e^-}$ and $B$) and of the CBM (e.g., $\alpha$, $n_{CBM}$, and $\mathcal{R}$).
\end{itemize}

It is interesting that the very simplified conditions encountered in the short GRBs in the absence of a SN event, which characterize the long GRBs \citep{Yu2014}, have allowed definite progress in understanding some fundamental GRB properties, e.g., the correlation of high energy emission to the BH formation. They can be adapted to the case of long GRBs. The points summarized above go a long way towards reaching a better understanding of \familyone and \familytwo long GRBs \citep{Yu2014}, as well as of the BdHNe \citep{Ruffini2014}.
We are confident that GRB 140619B is one of the best examples of short GRBs obtained with the current space technology. We sincerely hope that the results of our research will lead to new missions with greater collecting area and time resolution in X- and gamma-rays.

\acknowledgements

We thank the referee for requesting additional observational support for our theoretical fireshell binary merger model. This has motivated us to improve the connection of our theoretical work with the observations, resulting in this new version of the manuscript. 
We are especially grateful to S. Campana and C.~L. Fryer for useful suggestions in improving some conceptual and observational arguments, and to L. Izzo for the detailed analysis of the high energy emission. ME, MK and FGO are supported by the Erasmus Mundus Joint Doctorate Program by grant Nos. 2012-1710, 2013-1471, and 2012-1710, respectively, from the EACEA of the European Commission. AVP and EZ acknowledge the support by the International Cooperation Program CAPES-ICRANet financed by CAPES-Brazilian Federal Agency for Support and Evaluation of Graduate Education within the Ministry of Education of Brazil. This work made use of data supplied by the UK \textit{Swift} Science Data Centre at the University of Leicester.

\end{document}